\documentclass{tlp}
\usepackage{graphicx}
\usepackage{rotating}
        
\usepackage{aopmath}
\usepackage{amssymb}
\usepackage{proof}
\usepackage{euscript}
\usepackage{amsfonts}
\usepackage{latexsym}
\usepackage{epic,eepic}

\newtheorem{definition}{Definition}[section]

\renewcommand{\theenumi}{\mbox {\rm ${\roman{enumi}}$}}

\newtheorem {remark}[definition]{Remark}

\newtheorem {es}[definition]{Example}
\newenvironment {example} {\begin{es}\rm}{\hspace*{\fill}$\Box$\end{es}}
\newcommand {\comment}[1]{}
\newcommand {\Line} {{\bf -----------------------------------------------------------------------------------------}}

\newcommand{\eqdef}{\ensuremath{\stackrel{\mbox{\upshape\tiny def}}{=}}}

\newcommand {\Nat} {{\mathbb N}}

\newcommand {\Union} {\cup}

\newcommand {\myunion} {\bigcup}
\newcommand {\amyunion} {\bigsqcup}
\newcommand {\diff} {\!\setminus\!}
\newcommand {\intersect} {\cap}
\newcommand {\eset} {\emptyset}

\newcommand {\comp} {\circ}
\newcommand {\tuple} [1] {\langle {#1} \rangle}
\newcommand {\couple} [2] {\langle {#1},{#2} \rangle}

\newcommand {\Eps} {\epsilon}
\newcommand {\Part} [1] {\wp({#1})}
\newcommand {\mlub} [2] {#1{\mbox {\footnotesize $\bullet$}} #2}
\newcommand {\asubseteq} {\sqsubseteq}
\newcommand {\Inf} {\infty}
\newcommand {\submult} {\preccurlyeq}
\newcommand {\supmult} {\succcurlyeq}

\newcommand {\fun} [2] {{#1}\rightarrow{#2}}
\newcommand {\funct} [3] {{#1}:\fun{#2}{#3}}
\newcommand {\st} {\ |\ }
\newcommand {\MS} [1] {{\cal M\calS}({\mbox {${#1}$}})}
\newcommand {\card} [1] {|{#1}|}
\newcommand {\scalar} {\cdot}

\newcommand {\SummL} [2]{{\sum_{\mbox{\footnotesize{${#1}$}}}{#2}}}

\newcommand {\mmgu} {{\mathit mgu}}
\newcommand {\Mmgu} [2] {\mmgu({#1},{#2})}
\newcommand {\restr} [2] {{#1}_{|{#2}}}
\newcommand {\fvar} [1] {\mathit{FV}({#1})}
\newcommand {\Dom} [1] {\mathit{Dom}({#1})}
\newcommand {\compos} [2] {{#1}\comp{#2}}
\newcommand {\substbind} [2] {{#1}\mapsto{#2}}
\newcommand {\slub} {\uparrow}

\newcommand {\foral} [2] {\forall {#1}_{.}{#2}}
\newcommand {\foralx} [1] {\foral{x}{#1}}

\newcommand {\with} {\,\&\,}
\newcommand {\lolli} {\mathbin{-\hspace{-0.70mm}\circ}}
\newcommand {\lollo} {\mathbin{\circ\hspace{-0.70mm}-}}

\newcommand{\para}{\rotatebox[origin=c]{180}{\,\ensuremath\&\,}}

\newcommand {\tensor} {\otimes}
\newcommand {\anti} {\bot}
\newcommand {\all} {\top}
\newcommand {\one} { {\bf 1} }

\newcommand {\LOf} {LO${}_\forall$\,\,}

\newcommand {\LOfws} {LO${}_\forall$}

\newcommand {\nl} {\mbox{}\\}

\newcommand {\vsep} {\nl[\medskipamount]}
\newcommand {\hand} {\ \ \hbox{and}\ \ }

\newcommand {\sectionl} [2] {\section {#1} \label {#2}}
\newcommand {\subsectionl} [2] {\subsection {#1} \label {#2}}

\newcommand {\captionl} [2] {\caption {#1} \label {#2}}

\newcommand {\calA} {{\cal A}} 
\newcommand {\calB} {{\cal B}}
\newcommand {\calC} {{\cal C}}
\newcommand {\calD} {{\cal D}}

\newcommand {\calF} {{\cal F}}

\newcommand {\calI} {{\cal I}}

\newcommand {\calL} {{\cal L}}
\newcommand {\calM} {{\cal M}} 
\newcommand {\calN} {{\cal N}}
 
\newcommand {\calP} {{\cal P}}

\newcommand {\calS} {{\cal S}}

\newcommand {\calV} {{\cal V}}

\newcommand {\Sig} [1] {\mathit {Sig}_{#1}}
\newcommand {\SigP} {\Sig{P}}
\newcommand {\SigmaP} {{\Sigma_P}}

\newcommand {\Tsigma} {\mbox{$ T_{\Sigma} $}}
\newcommand {\TsigmaX} [1] {\mbox{$ T_{\Sigma}^{{#1}} $}}

\newcommand {\TsigV} [1] {\mbox{$ T_{#1}^\calV $}}
\newcommand {\TsigmaV} {\mbox{$ T_{\Sigma}^\calV $}}

\newcommand {\AsigV} [1] {\mbox{$ A_{#1}^\calV $}}
\newcommand {\AsigmaV} {\mbox{$ A_{\Sigma}^\calV $}}
\newcommand {\AsigmaPV} {\mbox{$ A_{\SigmaP}^\calV $}}

\newcommand {\gnd} [1] {\mathit {Gnd}({#1})}

\newcommand {\vrn} [1] {\mathit {Vrn}({#1})}
\newcommand {\INST} [1] {\mathit {Inst}_{#1}}
\newcommand {\instS} {\INST{\Sigma}}
\newcommand {\Inst} [2] {\INST{{#1}}({#2})}
\newcommand {\InstS} [1] {\Inst{\Sigma}{#1}}

\newcommand {\up} [1] {\mathit {Up}_{#1}}
\newcommand {\upS} {\up{\Sigma}}
\newcommand {\Up} [2] {\up{{#1}}({#2})}
\newcommand {\UpS} [1] {\Up{\Sigma}{#1}}

\newcommand {\os} [1] {\mathit {O}({#1})}
\newcommand {\osP} {\os{P}}

\newcommand {\fs} [1] {\mathit {F}({#1})}
\newcommand {\fsP} {\fs{P}}

\newcommand {\hb} [2] {\mathit {H\!B}_{#1}({#2})}
\newcommand {\hbS} [1] {\hb{\Sigma}{#1}}
\newcommand {\den} [1] {[\![{#1}]\!]}
\newcommand {\Fam} [1] {\{{#1}_\Sigma\}_{\Sigma\in\SigP}}
\newcommand {\iequiv} {\simeq}

\newcommand {\iclass} [1] {[{#1}]_\iequiv}
\newcommand {\Til} [3] {\lceil{#1}\rceil_{{\mbox{\tiny ${#2}$}}
                                          {\mbox{\tiny $\rightarrow$}}
                                          {\mbox{\tiny ${#3}$}}}}
\newcommand {\til} [1] {\lceil{#1}\rceil}
\newcommand {\Xis} [2] {\xi_{{\mbox{\tiny ${#1}$}}
                             {\mbox{\tiny $\rightarrow$}}
                             {\mbox{\tiny ${#2}$}}}}
\newcommand {\xis} {\xi}

\newcommand {\entailm} {\sqsubseteq^m}

\newcommand {\entailSet} {\sqsubseteq^s}

\newcommand {\ded}{\vdash}

\newcommand {\dedlo} [3] {{#1}\ded\!\!_{#2}\,{#3}}
\newcommand {\dedLO} [2] {{#1}\ded{#2}}
\newcommand {\dedLOo} [2] {{#1}\ded_\one{#2}}

\newcommand {\dedloP} [2] {\dedlo{P}{#1}{#2}}

\newcommand {\dedloPS} [1] {\dedlo{P}{\Sigma}{#1}}
\newcommand {\dedloPSc} [1] {\dedlo{P}{\Sigma,c}{#1}}
\newcommand {\dedloPScd} [1] {\dedlo{P}{\Sigma,c,d}{#1}}

\newcommand {\separ} {\,\mbox {\footnotesize
                      $\stackrel{\blacktriangleright}{}$}\,}
\newcommand {\satP} {\models}

\newcommand {\sat} [1] {\models_{#1}}
\newcommand {\satS} {\sat{\Sigma}}
\newcommand {\asat} [1] {\Vdash_{#1}}
\newcommand {\asatS} {\asat{\Sigma}}
\newcommand {\val} [4] {{#1} \sat{#2} {#3}\separ{#4}}
\newcommand {\valS} [3] {\val{#1}{\Sigma}{#2}{#3}}
\newcommand {\aval} [5] {{#1} \asat{#2} {#3}\separ{#4}\separ{#5}}
\newcommand {\avalS} [4] {\aval{#1}{\Sigma}{#2}{#3}{#4}}

\newcommand {\Tp} {T_P}

\newcommand {\SP} {S_P}
\newcommand {\itp} [1] {\Tp\!\!\uparrow_{#1}}
\newcommand {\isp} [1] {\SP\!\!\uparrow_{#1}}
\newcommand {\Itp} [2] {\itp{{#1}}\!\!({#2})}

\newcommand {\Un} [2] {\myunion_{{#1}=1}^{#2}}

\newcommand {\aUn} [2] {\amyunion_{{#1}=1}^{#2}}
\newcommand {\alub} [2] {{#1}\amyunion{#2}}

\newcommand {\ms}[1] {\widehat{#1}}

\newcommand {\lfp} [1] {{lfp}({#1})}

\newcommand {\SymbF} [1] {\mathit {\calF\!\!_{sym}} ({#1})}

\newcommand {\formula} [1] {{\bf {#1}}}

\newcommand {\Pre} {Pre}

\newcommand {\Pres} {Pre^*}

\newcommand {\DATAC} {{\textsf{\setbox0\hbox{T}da\raise.1\wd0\hbox to.4\wd0
                      {\hss T\hss}ac}}}

\title[Model Checking Linear Logic Specifications]
{Model Checking Linear Logic Specifications}
\author[M. Bozzano, G. Delzanno and M. Martelli]
{MARCO BOZZANO${\mbox {}^{1,2}}$, GIORGIO DELZANNO${\mbox {}^{2}}$ and MAURIZIO
MARTELLI${\mbox {}^{2}}$\\                       
\begin{tabular}{c}
\begin{tabular}{c}
\\
${\mbox {}^{1}}$ITC-IRST\\
Via Sommarive 18, Povo, 38050 Trento, Italy\\
\email{bozzano@irst.itc.it}
\end{tabular}
\bigskip\\
\begin{tabular}{c}
${\mbox {}^{2}}$Dipartimento di Informatica e Scienze dell'Informazione\\
Universit\`a di Genova\\
Via Dodecaneso 35, 16146 Genova - Italy \\
\email{\{giorgio,martelli\}@disi.unige.it}
\end{tabular}
\end{tabular}
}

\pagerange{\pageref{firstpage}--\pageref{lastpage}}

\volume{}
\jdate{}
\pubyear{}

\begin{document}

\label{firstpage}

\maketitle

\begin{abstract}
The overall goal of this paper is to investigate the theoretical
foundations of {\em algorithmic} verification techniques for
 {\em first order linear logic} specifications. 
The fragment of linear logic we consider in this paper is based on the linear 
logic programming language called LO \cite{AP90} enriched with {\em universally
quantified goal formulas}. Although LO was originally introduced
as a theoretical foundation for extensions of {\em logic
programming} languages, it can also be viewed as a very general
language to specify a wide range of {\em infinite-state concurrent systems}
 \cite{And92,Cer95}.

Our approach is based on the relation between {\em backward
reachability} and {\em provability} highlighted in our previous
work on {\em propositional} LO programs \cite{BDM02a}. Following this
line of research, we define here a general framework for the {\em
bottom-up} evaluation of {\em first order} linear logic specifications.
The evaluation procedure is based on an effective fixpoint
operator working on a symbolic representation of infinite
collections of first order linear logic formulas. The theory of
well quasi-orderings \cite{ACJT96,FS01} can be used to provide sufficient conditions
for the termination of the evaluation of non trivial fragments of
first order linear logic.
\end{abstract}


\section{Introduction}
\label{introsec}

The algorithmic techniques for the analysis of Petri Nets are based on
very well consolidated theoretical foundations
\cite{EM00,KM69,May84,EFM99,Fin93,STC98}.  However, several interesting
problems, e.g., the {\em coverability} problem,
become undecidable when considering
specification languages more expressive than basic Petri Nets.  In
this setting, {\em validation} of complex specifications is often
performed through {\em simulation} and {\em testing}, i.e., by
``executing'' the specification looking for design errors, e.g., as in
the methodology based on the construction of the reachability graph of
Colored Petri Nets \cite{Jen97}.  In order to study algorithmic
techniques for the analysis of a vast range of concurrency models it
is important to find a uniform framework to reason about their
characteristic features.

In our approach we will adopt {\em linear logic} \cite{Gir87}
as a unified  logical framework for concurrency.
Linear logic provides a logical characterization of concepts and
mechanisms peculiar of concurrency like {\em locality}, {\em recursion}, and
{\em non determinism} in the definition of a process \cite{AP90,KY95,MM91};
communication via {\em synchronization} and {\em value passing} \cite{Cer95,Mil93};
{\em internal state and updates to its current value} \cite{AP90,Mil96};
and {\em generation of fresh names} \cite{CDLMS99,Mil93}.
{\em Provability} in fragments of linear logic can be used as a
formal tool to reason about behavioral aspects of the {\em concurrent systems}
\cite{BDM02a,MMP96}.

The overall goal of this paper is to investigate the theoretical
foundations of {\em algorithmic} verification techniques for specifications
based on {\em first order linear logic}.
The fragment we consider in this paper is based on the linear
logic programming language called LO \cite{AP90} enriched with
{\em universally quantified goal formulas}. Apart from being a logic programming
language, the appealing feature of LO  is that it can also be viewed as a 
rich {\em specification} language for concurrent systems:
\begin{itemize}
\item Specification languages like Petri Nets and multiset
rewriting over first order atomic formulas 
can be naturally embedded into {\em propositional LO}
(see, e.g., \cite{Cer95,BDM02a}). 
\item {\em First order LO specifications} can be used to specify the internal state
of processes with structured data represented as terms, thus
enlarging the class of systems that can be formally specified in
the logic. In this context universal quantification in goal
formulas has several interesting interpretations: it can be viewed
either as a sort of {\em hiding} operator in the style of
$\pi$-calculus \cite{Mil93}, or as a mechanism to generate {\em
fresh names} as in \cite{CDLMS99}.
\end{itemize}
Before discussing in more details the technical contributions of our
work, we will briefly illustrate the connection
between Petri Nets and linear logic, and between reachability and
provability in the corresponding formal settings. The bridge between
the two paradigms is the {\em proofs as computations} interpretation
of linear logic proposed in \cite{And92} and in \cite{Mil96}.
\paragraph{Linear Logic and Concurrency}
A Petri Net can be represented by means of a multiset-rewriting
system over a finite alphabet, say $p,q,r,\ldots$, of {\em place} names.
One possible way of expressing multiset rewrite rules in linear logic is based on the following
idea. The connective $\para$ (multiplicative disjunction) is interpreted as a
multiset constructor, whereas the connective $\lollo$ (reversed linear implication) is
interpreted as the rewrite relation.
Both connectives are allowed in the LO fragment.
For instance, as shown in \cite{Cer95} the LO clause
$$
p\para q~\lollo~p \para p \para q \para  t
$$
can be viewed as a Petri Net {\em transition} that removes a token from places
$p$ and $q$ and puts two tokens in place $p$, one in $q$, and one in $t$.
According to the proofs as computations interpretation \cite{And92},
a {\em top-down} derivation in linear logic consists of a goal-directed sequence of rule
applications. If we look at the initial goal as a multiset of atomic formulas (places)
representing the initial marking of a Petri Net, then each application of an LO clause
like the one illustrated above ({\em backchaining} in the terminology of \cite{And92}) simulates
the firing of a Petri Net transition at the corresponding marking. Furthermore, the overall
top-down derivation corresponds to one of the possible executions of the net,
leading from the initial marking to one of the target states.

Thanks to the presence of other connectives, LO supports more sophisticated mechanisms
than the ones available in simple Petri Nets.
For instance, in \cite{AP90} Andreoli and Pareschi use LO clauses with
occurrences of $\para$  and $\with$ (additive conjunction) in their {\em body}
to express what they called {\em external} and {\em internal} concurrency.
Additive conjunction can be used, in fact, to simulate independent threads of execution
running in parallel.

In our previous work \cite{BDM02a}, we made a first attempt to connect
techniques used for the validation of Petri Nets with evaluation strategies
of LO programs.
Specifically, in \cite{BDM02a} we defined an effective procedure to compute the
set of  linear logic {\em goals} (multisets of atomic formulas)
that are consequences of a given propositional program, i.e.,
a ``bottom-up'' evaluation procedure for {\em propositional} LO programs.
Our construction is based on the {\em backward reachability}
algorithm of \cite{ACJT96} used to decide the so called
{\em control state  reachability problem} of Petri Nets
(i.e., the problem of deciding if a given set of {\em upward closed} configurations are
reachable from an initial one).
The algorithm works as follows.
Starting from a set of {\em target states},
the algorithm computes symbolically the transitive closure of
the {\em predecessor} relation (i.e., the transition relation read backwards)
of the Petri Net taken into consideration.
The algorithm is used to check safety properties:
if the algorithm is executed starting from the set of {\em unsafe states},
then the corresponding safety property holds if and only if the
initial marking is not in the resulting fixpoint.

In order to illustrate the connection between backward reachability for
Petri Nets and provability in LO, we first observe that LO program clauses
of the form
$$
p\para q\para q~\lollo~\all
$$
succeed in any context containing {\em at least} one occurrence of $p$ and two occurrences
of $q$.
In other words they can be used to symbolically represent sets of markings
that are closed upwards with respect to the multiset  inclusion relation.
Now, suppose we represent a Petri Net via an LO program $P$ and
the set of {\em target states} using a collection $T$ of LO program clauses with $\all$ in
the body.
Then, the set of {\em facts} (i.e., multisets of atomic formulas)
that are {\em logical consequences} of the LO program $P\cup T$
will represent the set of markings that are {\em backward reachable}
from the target states.

The algorithm we presented in \cite{BDM02a} is based on this idea, and it
extends the backward reachability algorithm for Petri Nets of \cite{ACJT96}
to the more general case of propositional LO programs (i.e., with nested conjunctive
and disjunctive goals).
\paragraph{First Order Linear Logic}
By lifting the logic language to {\em first order},
the resulting specification language becomes much more interesting and flexible than
basic Petri Nets.
In the extended setting, the logic representation of processes
can be enriched with a notion of {\em internal state} and with communication mechanisms
in which {\em values} can be passed between different processes.
As an example, the following LO clause
$$
idle(Y)\para p(alice,wait,stored(Y))~\lollo~p(alice,use,stored(Y))
$$
can be interpreted as a transaction of a protocol during which the {\em process} named Alice
(currently knowing $Y$) synchronizes with a monitor controlling the resource $Y$, 
checks that the monitor is {\em idle} and then enters the critical section in which she uses
the resource $Y$.
By instantiating the free variables occurring in such a rule, we obtain a family of transition
rules that depend on the domain used to define the content of messages.
In this setting the universal quantification in goal formulas can be used to generate
{\em fresh  values}, as in the following rule:
$$
init \lollo~\foralx{~idle(x)~\para~init}
$$
Intuitively, the demon process $init$ creates new resources labeled with fresh 
identifiers.


The above illustrated connection between {\em provability} and {\em
reachability} immediately gives us a well-founded manner of extending
the algorithmic techniques used for the analysis of Petri Nets to the
general case of first order linear logic specifications.
%
\paragraph{Our Contribution}
The conceptual and technical contributions of our work can be summarized
as follows.
\begin{itemize}
\item[(1)] Combining ideas coming from the semantics of logic programming \cite{BGLM94,FLMP93} 
and from symbolic model checking for infinite state systems \cite{ACJT96,FS01},
in this paper we present the theoretical foundations for the definition of a procedure
for the {\em bottom-up} evaluation of first order LO programs with universally quantified goals.
By working in the general setting of linear logic, we obtain a framework that can
be applied to other specification languages for concurrent systems like 
{\em multiset rewriting over first order atomic formulas} \cite{CDLMS99}.

The bottom-up evaluation procedure can also be viewed as a {\em fixpoint}
semantics that allows us to compute the set of all goals that are
linear logical consequences of a given (extended) LO program.
The fixpoint semantics is based on an {\em effective} fixpoint operator and on a 
{\em symbolic}
and {\em finite} representation of an {\em  infinite} collection of first order
provable LO goals. As previously mentioned, the possible infiniteness of the set of
provable goals is due to LO program clauses with the constant $\all$, which represent 
sets of goals which
are upward-closed with respect to the multiset inclusion relation.
The symbolic representation is therefore crucial when trying to prove properties of
infinite systems like {\em parameterized} systems, i.e., systems in which the  number of
individual processes is left as a parameter of the specification
(e.g., mutual exclusion protocols for {\em multi-agent} systems \cite{Boz02}).
Intuitively, such a representation is obtained by restricting our attention to logical 
consequences represented via multisets of first order atomic formulas.
As an example, the formula
$$
p(A,use,stored(X))~\para~p(B,use,stored(X))~\lollo~\all
$$
can be used to denote all multisets of {\em ground} atomic formulas
{\em containing} an instance of the clause head. As the constant
$\all$ is provable in any context, in the previous example we obtain a
{\em symbolic representation} of the infinite set of {\em unsafe
states} generated by the following minimal violation of mutual
exclusion for a generic resource represented via the shared variable
$X$: {\em at least two different processes are in their critical
section using a shared resource}.
\item[(2)] Besides the connection with verification of concurrent
systems, the new fixpoint semantics for first order LO programs
represents an alternative to the traditional top-down execution of
linear logic programs studied in the literature \cite{And92}.  Thus,
also from the point-of-view of logic programming, we extend the
applicability of our previous work \cite{BDM02a} (that was restricted
to the propositional case) towards more interesting classes of linear
logic programs.
\item[(3)] The termination of the fixpoint computation cannot be
guaranteed in general; first order LO programs are in fact Turing
complete.  However, we present here sufficient conditions under which
we can compute a {\em symbolic representation of all logical
consequences} of a non trivial first order fragment of LO with
universal quantification in goal formulas.  As a direct consequence of
this result, we obtain that provability is decidable in the considered
fragment.  To our knowledge, this result uncovers a new decidable
fragment of first order linear logic.  The fragment taken into
consideration is not only interesting from a theoretical point of
view, but also as a possible abstract model for ``processes'' with
identifiers or local values.
\end{itemize}

Though the emphasis of this work is on the theoretical grounds of our
method, we will illustrate the practical use of our framework with the
help of a verification problem for a {\em mutual exclusion protocol}
defined for a concurrent system which is parametric in the number of
{\em clients}, {\em resources}, and related {\em monitors}.  Other
practical applications of this method are currently under
investigation.  Preliminary results in this direction are shown in the
PhD thesis of Marco Bozzano \cite{Boz02}.

Finally, we remark that a very preliminary version of this work appeared
in the proceedings of FLOPS 2001 \cite{BDM01a}.

\subsection{Outline of the Paper}
The terminology and some notations used in the paper are presented in
Appendix \ref{prelimsec}.  To improve the readability of the paper,
the proofs of some lemmas are given in Appendix \ref{proofsapp}.  In
Section \ref{relatedworks}, we will discuss related works.  In Section
\ref{losec} we will recall the main definitions of the fragment LO of
\cite{AP90}, presented here with universal quantification in goal
formulas.  In order to illustrate the use of LO as a specification
logic for concurrent systems, in the same section we will briefly
describe how {\em multiset rewriting} (extended with quantification)
can be embedded into LO.  This connection represents a natural entry
point into the world of concurrency.  In fact, the relationship
between multiset rewriting, (Colored) Petri Nets, and process calculi
has been extensively studied in the literature (see e.g.,
\cite{Cer95,Far99,Far00,Mes92,MM91}).  Finally, we will present an example
of use of LO as a specification language for concurrent systems, and
discuss the relationships between (bottom-up) LO provability and
verification techniques based on (infinite-state) model checking.  In
Section \ref{fobottomupsec}, we will introduce a {\em non effective}
fixpoint semantics for linear logic programs. To simplify the
manipulation of non ground terms, we will first lift the top-down
(proof theoretical) semantics of LO to the non ground level, by
introducing a new proof system in which sequents may have formulas
with free variables.  In Section \ref{foeffectivesec}, we will
introduce a general framework for the bottom-up evaluation of LO
programs. The bottom-up procedure is based on a finite representation
of infinite sets of logical consequences, and on an effective fixpoint
operator working on sets of symbolic representations. The bottom-up
procedure can be seen as a symbolic version of the semantics presented
in Section \ref{fobottomupsec}. The reason for introducing two
different semantic definitions is to ease the proof of soundness and
completeness, which is split into the proof of equivalence of the
effective semantics with respect to the non-effective one, and the
proof of equivalence of the non-effective semantics with respect to
the operational one.  In Section \ref{foterminatsec}, we will
investigate sufficient conditions for the termination of the bottom-up
evaluation.  In Section \ref{examplessec}, we will discuss the
possible application of the resulting method as a verification
procedure for {\em infinite-state parameterized systems}.  In Section
\ref{reachability}, we will address possible future directions of
research.  In Section \ref{conclusions}, we will address some
conclusions.
\section{Related Works}
\label{relatedworks}
To our knowledge, our work is the first attempt
to connect algorithmic techniques used in symbolic model checking
with declarative and operational aspects of {\em first order} linear logic programming.
In \cite{BDM02a}, we have considered the relation
between {\em propositional} LO and Petri Nets.
Specifically, in \cite{BDM02a} we have shown that the bottom-up semantics 
is computable for propositional LO programs 
(because of the relationship of this  problem with the coverability problem of  Petri Nets).
Furthermore, in \cite{BDM02a} we have shown that the bottom-up  
evaluation of propositional LO programs enriched with the constant $\one$ 
is not computable in a finite number of steps (otherwise
one could decide the equivalence problem for Petri Nets).

We point out here that an original contribution of the paper 
consists in extending the construction we used for proving the computability 
of the bottom-up construction of propositional LO programs to {\em first order}
LO specifications. This way, we have established a link with more complex 
models of concurrency. 
Clearly, in the first order case provability becomes undecidable.
In the paper we present a non trivial special case of first order LO programs 
in which the bottom-up semantics is still computable.
Extending the bottom-up evaluation to LO programs enriched with
the constant $\one$ is a possible future direction of research (see Section
\ref{reachability} for a discussion).

In \cite{HW98}, Harland and Winikoff  present an abstract deductive system
for bottom-up evaluation of linear logic programs.
The left introduction plus weakening and cut rules are used to compute
the logical consequences of a given formula.
Though the framework is given for a more general fragment than {LO},
it does not provide an {\em effective} procedure to evaluate
programs.
In \cite{APC97}, Andreoli, Pareschi and  Castagnetti define an improved
{\em top-down} strategy for {\em propositional} LO based on the
Karp-Miller's covering graph of Petri Nets, i.e., a {\em forward} exploration
with accelerations.

The relation between Rewriting, (Colored) Petri Nets and Linear Logic has been
investigated in previous works like \cite{Cer94b,Cer95,EW90,Mes92,MM91}.
Our point-of-view is based on the proofs as computations metaphor proposed in 
\cite{AP90,And92,Mil96}, whereas
our connection with models for concurrency is inspired to works in this field like
\cite{Cer94b,Cer95,DM01,KY94,Mil93,Mil96}. As an example, in \cite{Cer94b,Cer95},
Cervesato shows how to encode Petri Nets in different fragments of linear logic like
{LO}, Lolli \cite{HM89}, and Forum \cite{Mil96} exploiting the different
features of these languages. Algorithmic aspects for verification of properties of the
resulting linear logic specifications are not considered in the works mentioned above.
In \cite{Far99,Far00}, Farwer presents a possible encoding of Colored Petri Nets in Linear Logic
and proposes a combination of the two formalisms that could be used to model object systems.

The problem of the {\em decidability} of provability in fragments of linear logic has been
investigated in several works in recent years \cite{Lin95,LMSS92,LSce94}.
Specifically, in \cite{Kop95}, Kopylov has shown that the full {\em propositional} linear
{\em affine}  logic containing all the multiplicatives, additives, exponentials,
and constants is decidable. Affine logic can be viewed as linear logic with the
{\em weakening rule}. Propositional LO belongs to such a sub-structural logic.
Provability in full first order linear logic is undecidable as shown by Girard's
translation of first order logic into first order linear logic \cite{Gir87}.
The same holds for first order  affine logic (Girard's encoding can also be viewed as
an encoding into affine logic \cite{Lin95}).
First order linear logic {\em without modalities}, i.e.,
without the possibility of re-using formulas, is decidable \cite{LSce94}.
In \cite{CDLMS99}, Cervesato et al. use a formalism based on
multiset-rewriting and existential quantification that can be embedded into our fragment
of linear logic to specify protocol rules and actions of intruders.
In \cite{DLMS99}, it is shown that reachability in
multiset rewriting with existential quantification is undecidable by a reduction
from Datalog with quantification in goal formulas. The fragment they consider however
is much more general than the {\em monadic} fragment of \LOfws.
Monadic \LOf can be viewed as a fragment of first order linear affine logic
with restricted occurrences of the exponentials (program clauses are re-usable)
and severe restrictions on the form of atomic formulas.
We are not aware of previous results on similar fragments.
\sectionl{The Logic Programming Language LO}{losec}
LO \cite{AP91a} is a logic programming language based on a fragment of
LinLog \cite{And92}.
Its mathematical foundations lie on a proof-theoretical presentation of
a fragment of linear logic defined over the linear connectives
$\lolli$ ({\em linear implication}, we use the reversed notation
$H\lollo G$ for $G\lolli H$), $\with$ ({\em additive conjunction}),
$\para$ ({\em multiplicative disjunction}), and the constant $\all$
({\em additive identity}). In this section we present the proof-theoretical
semantics, corresponding to the usual {\em top-down}
operational semantics for traditional logic programming languages,
for an extension of LO. First of all, we consider a slight extension of LO
which admits the constant $\anti$ in goals and clause heads. More importantly,
we allow the universal quantifier to appear, possibly nested, in goals.
This extension is inspired by
{\em multiset rewriting with universal quantification} \cite{CDLMS99}.
The resulting language will be called \LOf hereafter.
Following \cite{AP91a}, we give the following definitions.
\begin{definition}[Atomic Formulas]
Let $\Sigma$ be a signature with predicates
including a set of constant and function symbols
$\calL$ and a set of predicate symbols $\calP$, and let $\calV$ be a
denumerable set of variables. An atomic formula over $\Sigma$ and $\calV$
has the form $p(t_1,\ldots,t_n)$ (with $n\geq 0$),
where $p\in\calP$ and $t_1,\ldots,t_n$
are (non ground) terms in $\TsigmaX{\calV}$.
We denote the set of such atomic formulas as $A_\Sigma^\calV$.
\end{definition}
We are now ready to define \LOf programs. The class of $\formula{D}$-formulas
correspond to multiple-headed program clauses, whereas $\formula{G}$-formulas
correspond to {\em goals} to be evaluated in a given program.
\begin{definition}[\LOf programs]
Let $\Sigma$ be a signature with predicates and $\calV$ a denumerable set
of variables. The classes of $\formula{G}$-formulas (goal formulas),
$\formula{H}$-formulas (head formulas), and $\formula{D}$-formulas
(program clauses) over $\Sigma$ and $\calV$ are defined by the following
grammar:
$$
\begin{array}{l}
  \formula{G}\ ::=\ \formula{G}\ \para\ \formula{G}\ \ |\ \formula{G}\ \with\
    \formula{G}\ \ |\ \ \foralx{\formula{G}}\ \ |\ \
    \formula{A}\ \ |\ \ \all\ \ |\ \ \anti\\
  [\smallskipamount]
  \formula{H}\ ::=\ \formula{A}\para\ \ldots\para\ \formula{A}\ \ |\ \
    \anti\\
  [\smallskipamount]
  \formula{D}\ ::=\ \formula{H}\ \lollo\ \formula{G}\ \ |\ \
  \formula{D}\ \with\ \formula{D}\ \ |\ \ \foralx{\formula{D}}
\end{array}$$
where $\formula{A}$ stands for an atomic formula over $\Sigma$ and $\calV$.
An \LOf program over $\Sigma$ and $\calV$ is a $\formula{D}$-formula
over $\Sigma$ and $\calV$.
A multiset of goal formulas will be called a {\em context} hereafter.
\end{definition}
\begin{remark}\rm
\label{programremark}
Given an \LOf program $P$, in the rest of the paper
we often find it convenient to view $P$ as the {\em set} of clauses
$D_1,\ldots,D_n$. Every {\em program clause} $D_i$ has the form 
$\forall\,(H\lollo G)$ standing for $\forall x_1\ldots x_k\ldotp(H\lollo G)$, 
where $\fvar{H\lollo G}=\{x_1,\ldots,x_k\}$.
\\
Formally, this is justified by the following logical equivalences \cite{Gir87}:
$$
\begin{array}{c}
!(D_1\with D_2)~\equiv~!D_1~\tensor~!D_2\\
\foralx{(D_1\with D_2)}~\equiv~~\foralx{D_1}~\with~\foralx{D_2}
\end{array}
$$ 
\end{remark}
For the sake of simplicity, in the following we usually omit the universal
quantifier in $\formula{D}$-formulas, i.e., we consider free variables as
being {\em implicitly} universally quantified.
\begin{definition}[\LOf Sequents]
Let $\Sigma$ be a signature with predicates and $\calV$ a denumerable set
of variables. An \LOf sequent has the form $\dedloP{\Sigma'}{G_1,\ldots,G_k}$,
where $P$ is an \LOf program over $\Sigma$ and $\calV$,
$G_1,\ldots,G_k$ is a context (i.e., a multiset of goals)
over $\Sigma$ and $\calV$, and $\Sigma'$ is a signature such that
$\Sigma\subseteq\Sigma'$.
\end{definition}
According to Remark \ref{programremark}, structural rules 
({\em exchange}, {\em weakening} and {\em contraction}) are allowed
 on the left-hand side, while on the
right-hand side only the rule of exchange is allowed
(for the fragment under consideration, it turns out that the
rule of weakening is admissible, while contraction is forbidden).
We now define provability in \LOfws.
\begin{definition}[Ground Instances]
\label{logroundinstdef}
Let $\Sigma$ be a signature with predicates and $\calV$ a denumerable set
of variables. Given an \LOf program $P$ over $\Sigma$ and $\calV$,
the set of ground instances of $P$, denoted $\gnd{P}$, is defined as follows:
$\gnd{P}\eqdef\{(H\lollo G)\,\theta\st\forall\,(H\lollo G)\in P\}$,
where $\theta$ is a grounding substitution for $H\lollo G$ (i.e., it maps
variables in $\fvar{H\lollo G}$ to ground terms in $\Tsigma$).
\end{definition}
The execution of a multiset of $\formula{G}$-formulas $G_1,\ldots,G_k$
in $P$ corresponds to a {\em goal-driven} proof for the sequent
$\dedloP{\Sigma}{G_1,\ldots,G_k}$.
According to this view, the operational semantics
of \LOf
is given via the {\em uniform} ({\em focusing}) \cite{And92} proof system
presented in Figure \ref{system_for_lof}, where
$P$ is a set of clauses, ${\cal A}$ is a multiset of atomic formulas,
and $\Delta$ is a multiset of $\formula{G}$-formulas.
We have used the notation $\ms{H}$, where $H$ is a
linear disjunction of atomic formulas $A_1\para\ldots\para A_n$,
to denote the multiset $A_1,\ldots,A_n$ (by convention, $\ms{\anti}=\Eps$,
where $\Eps$ is the empty multiset).
\begin{definition}[\LOf provability]
Let $\Sigma$ be a signature with predicates and $\calV$ a denumerable set
of variables. Given an \LOf program $P$ and a goal $G$,
over $\Sigma$ and $\calV$, we say that $G$ is provable from $P$ if
there exists a proof tree, built over the proof system of Figure
\ref{system_for_lof}, with root $\dedloPS{G}$, and such that every branch
is terminated with an instance of the $\all_r$ axiom.
\end{definition}
The concept of {\em uniformity} applied to LO
requires that the right rules $\all_r$, $\para_r$, $\with_r$, $\anti_r$,
$\forall_r$ have priority
over $bc$, i.e., $bc$ is applied only when the right-hand side of a sequent
is a multiset of {\em atomic} formulas (as suggested by the notation $\calA$
in Figure \ref{system_for_lof}).
The proof system of Figure \ref{system_for_lof} is a specialization of
more general uniform proof systems for linear logic like Andreoli's
focusing proofs \cite{And92} and Forum \cite{Mil96}.
\begin{figure*}
$$
\begin{array}{c}
\infer[\all_r]
{\dedloPS{\all,\Delta}}
{}
\ \ \ \
\infer[\para_r]
{\dedloPS{G_1\para G_2,\Delta}}
{\dedloPS{G_1,G_2,\Delta}}
\ \ \ \
\infer[\with_r]
{\dedloPS{G_1\with G_2,\Delta}}
{\dedloPS{G_1,\Delta} &
 \dedloPS{G_2,\Delta}}
\\
\\
\infer[\anti_r]
{\dedloPS{\anti,\Delta}}
{\dedloPS{\Delta}}
\ \ \ \
\infer[\forall_r\ \ (c\not\in\Sigma)]
{\dedloPS{\foralx{G},\Delta}}
{\dedloP{\Sigma,c}{G[c/x],\Delta}}
\ \ \
\infer[bc\ \ (H\lollo G\ \in\gnd{P})]
{\dedloPS{\ms{H},\calA}}
{\dedloPS{G,\calA}}
\end{array}
$$
\caption{A proof system for \LOf}
\label{system_for_lof}
\end{figure*}
Rule {\em bc} is analogous to a backchaining (resolution) step in
traditional logic programming languages.
Note that according to the concept of resolution explained above,
{\em bc} can be executed only if the right-hand side of
the current \LOf sequent consists of atomic formulas.
As an instance of rule $bc$, we get the following proof fragment, which
deals with the case of clauses with empty head:
$$
\begin{array}{c}
 \infer[bc] {\dedloPS{\calA}}
{\infer*{\dedloPS{\calA,G}}{}}
\medskip\\
provided\ \anti\lollo G\in\gnd{P}
\end{array}
$$
Given that clauses with empty head are always applicable in atomic
{\em contexts},
the degree of non-determinism they introduce in proof search is usually
considered unacceptable \cite{Mil96} and in particular they are forbidden in
the original presentation of LO \cite{AP91a}. However, the computational
model we are interested in, i.e., bottom-up evaluation, does not suffer
this drawback. Clauses with empty head often allow more flexible
specifications.

LO clauses having the form $H\lollo\bot$ simply remove 
the resources associated with $H$ from the right-hand side of the current sequent
($H$ is rewritten into the empty multiset).
On the contrary, LO clauses having the form $H\lollo\all$ can be viewed as {\em termination} 
rules.
In fact, when a backchaining step over such a clause is possible,
we get a {\em successful} (branch of a) computation,
independently of the current {\em context} $\calA$,
as shown in the following proof scheme:
$$
\begin{array}{c}
\infer[bc]
{\dedloPS{\ms{H},\calA}}
{\infer[\all_r]
  {\dedloPS{\all,\calA}}{}
}
\medskip\\
provided\ H\lollo\all\in\gnd{P}
\end{array}
$$
This observation is formally stated in the following proposition
(we recall that $\submult$ is the multiset inclusion relation).
\begin{proposition}[Admissibility of the Weakening Rule]
\label{foaffine}
Given an \LOf program $P$ and two multisets of goals $\Delta,\Delta'$
such that $\Delta\submult\Delta'$, if
$\dedloPS{\Delta}$ then $\dedloPS{\Delta'}$.
\end{proposition}
\begin{proof}
By simple induction on the structure of \LOf proofs.
\end{proof}
Admissibility of the weakening rule makes \LOf an
{\em affine} fragment of linear logic \cite{Kop95}.
Note that all structural rules are admissible on the left hand side
(i.e., on the program part) of \LOf sequents.

Finally, rule $\forall_r$ can be used to {\em dynamically} introduce new
{\em names} during the computation. The initial signature $\Sigma$ must
contain at least the constant, function, and predicate symbols of a given
program $P$, and it can dynamically grow thanks to rule $\forall_r$.

\begin{remark}\rm
\label{extrusion}
Particular attention must be paid to the constants introduced in a derivation.
They cannot be extruded from the scope of the corresponding universal
quantifier. For this reason, every time rule $\forall_r$ is applied, a new
constant $c$ is added to the current signature, and the resulting goal is
proved in the new signature.
The idea is that all terms appearing on the right-hand
side of a sequent are implicitly assumed to range over the relevant signature.
This behavior is standard in logic programming languages \cite{MNPS91}.
\end{remark}
\begin{example}
\label{lofex}
Let $\Sigma$ be a signature with a constant symbol $a$, a function symbol $f$
and predicate symbols $p,q,r,s$.
Let $\calV$ be a denumerable set of variables, and
$u,v,w,\ldots\in\calV$. Let $P$ be the program
$$\begin{array}{l}
  1\ldotp\ \ r(w)\lollo q(f(w))\para s(w)\\
  [\smallskipamount]
  2\ldotp\ \ s(z)\lollo\foralx{p(f(x))}\\
  [\smallskipamount]
  3\ldotp\ \ \anti\lollo q(u)\with r(v)\\
  [\smallskipamount]
  4\ldotp\ \ p(x)\para q(x)\lollo\all\\
\end{array}$$
The goal $s(a)$ is provable from $P$. The corresponding proof is shown
in Figure \ref{LOfprooffig} (where we have denoted by $bc^{(i)}$ the
application of the backchaining rule over clause number $i$ of $P$).
\begin{figure*}
$$
   \infer[bc^{(2)}]{\dedloPS{s(a)}}
  {\infer[\forall_r]{\dedloPS{\foralx{p(f(x))}}}
  {\infer[bc^{(3)}]{\dedloPSc{p(f(c))}}
  {\infer[\with_r]{\dedloPSc{p(f(c)),q(f(c))\with r(c)}}
  {\infer[bc^{(4)}]{\dedloPSc{p(f(c)),q(f(c))}}
  {\infer[\all_r]{\dedloPSc{\all}}{}}
    &
   \infer[bc^{(1)}]{\dedloPSc{p(f(c)),r(c)}}
  {\infer[\para_r]{\dedloPSc{p(f(c)),q(f(c))\para s(c)}}
  {\infer[bc^{(4)}]{\dedloPSc{p(f(c)),q(f(c)),s(c)}}
  {\infer[\all_r]{\dedloPSc{\all},s(c)}{}
  }}}}}}}
$$
\captionl{An example of \LOf proof}{LOfprooffig}
\end{figure*}
Note that the notion of {\em ground instance} is now relative to the current
signature. For instance, backchaining over clause 3 is possible because
the corresponding signature contains the constant $c$ (generated one level
below by the $\forall_r$ rule), and therefore
$\anti\lollo q(f(c))\with r(c)$ is a valid instance of clause 3.

\end{example}
\subsectionl{Simulating Multiset Rewriting over First Order Atoms}{unified}
In this section we will focus our attention on the relationship between
{\em multiset rewriting over first order atoms} and
{\em first order LO theories}.
We will conclude by showing how enriching logic theories with universal
quantification can provide a way to generate {\em new values}.

The connection between multiset rewriting systems over (first order) atomic
formulas and (first order) LO theories has been studied, e.g., 
in \cite{Cer94b,CDLMS99}.
In \cite{Cer94b} Cervesato presents different possible encodings of multiset
 rewriting (without function symbols) in linear logic. 
 Specifically, he first presents an encoding in the multiplicative fragment 
 of intuitionistic linear logic (MILL), 
 where multiplicative conjunction $\tensor$ (``tensor'') and linear implication are used as
  multiset constructor and rewrite relation, respectively.
  As an example, the formula $p \tensor q \lolli r \tensor s$ 
 represents a rewrite rule in which $p$ and $q$ are rewritten into $r$ and $s$ 
 ($\tensor$ denotes the ``tensor'').
  
As highlighted in Remark 5.12 of \cite{Cer94b} an equivalent encoding can 
be given by  choosing a fragment of classical linear logic contained 
in LO in which 
 multiplicative disjunction  and reverse linear implication are used as
  multiset constructor and rewrite relation, respectively.
 As an example, the formula $p \para q \lollo r \para s$ 
 represents a rewrite rule in which $p$ and $q$ are rewritten into $r$ and $s$.
  This is the encoding we will adopt in our work.
  
The duality of the two encodings is a consequence of the following property:
  	       $
	       (p \tensor q \lolli r \tensor s)^\bot \equiv 
	       p^\bot \para 
	       q^\bot \lollo  r^\bot 
	       \para ~s^\bot
	       $,
where $a^\bot$ is the linear logic negation of $a$.
Furthermore, it depends on  the way proofs are interpreted as computations, 
i.e., on whether ``rewrite rules'' are encoded as formulas that occur 
on the left- or on the right-hand side of a sequent.
  
In Section 5.2.2 of \cite{Cer94b} Cervesato also presents an encoding of Petri Nets 
  in LO that allows one to simulate the execution of a net using an LO top-down 
  derivation of the resulting program.
  In Section 5 of  \cite{CDLMS99} the encoding of multiset rewriting over first
  order atomic formulas (MSR) is
  extended to first order  MILL with existential quantifiers.
  Thanks to its logical nature,  the duality with the first order fragment
  of LO still holds. 

To illustrate the main ideas behind the interpretation of LO 
as multiset rewriting, let us first define the following class of LO formulas.
\begin{definition}\label{lorewrite}
We call {\em LO rewrite rule} any LO formula having the following form
  $$
  \forall(A_1\para\ldots\para A_n~\lollo~B_1\para\ldots\para B_m)
  $$
where $A_1,\ldots,A_n$ and $B_1,\ldots,B_m$ are atomic formulas over $\Sigma$ and $\calV$.
\end{definition}
As usual, the notation $\forall\,(H\lollo G)$ stands for the universal
quantification of clause $H\lollo G$ over its free variables.

LO formulas having the form depicted above can be interpreted as 
{\em multiset rewriting rules} in which rewriting can be performed 
only at the level of atomic formulas  as in the MSR framework 
defined in \cite{CDLMS99}.

Specifically, let $P$ be a set of LO rewrite rules
(as in Def. \ref{lorewrite}).
Now, consider a goal formula $G$ having the form
$C_1\para\ldots\para C_k$ where  $C_1,\ldots,C_k$ are {\em ground} atomic formulas over $\Sigma$.
It is easy to verify that any derivation starting from $\dedloPS{G}$
and built using LO proof rules amounts to a sequence of multisets rewriting steps,
where $\para$ is interpreted as multiset constructor.
\begin{example}
\label{msrex}
Let $\Sigma$ be a signature with two constant symbols $a$ and $b$, one
function symbol $f$ and two predicate symbols $p,q$.
Let $\calV$ be a denumerable set of variables and $x,y\in\calV$.
Let $P$ consists of the LO clause
$$
  \forall x,y\ldotp p(x)\para q(f(y)) \lollo p(f(x))\para q(y)\para q(f(x))
$$
and $G=p(a)\para p(b)\para q(f(b))$. Figure \ref{LOmsr} shows
one possible sequence of applications of the above clause that
starts from the sequent $\dedloPS{G}$ (we have underlined atomic
formulas selected in the application of the $bc$ rule).
\begin{figure*}
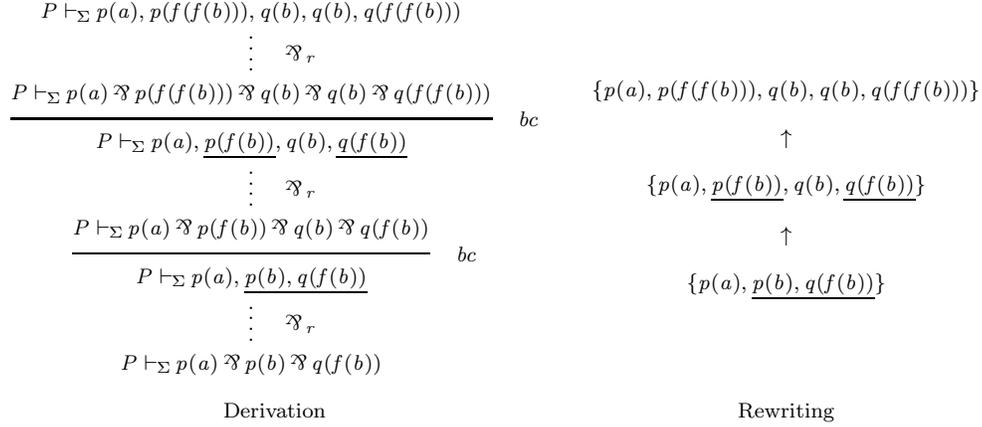

{\footnotesize
$$
\begin{array}{cc}
\begin{array}{c}
   \infer*[\para_r]{\dedloPS{p(a)\para p(b)\para q(f(b))}}
   {\infer[bc]{\dedloPS{p(a),{\underline{p(b),q(f(b))}}}}
   {\infer*[\para_r]{\dedloPS{p(a)\para p(f(b))\para q(b)\para q(f(b))}}
    {\infer[bc]{\dedloPS{p(a),\underline{p(f(b))},q(b),\underline{q(f(b))}}}
     {\infer*[\para_r]{\dedloPS{p(a)\para p(f(f(b)))\para q(b)\para q(b)\para q(f(f(b)))}}
      {\dedloPS{p(a),p(f(f(b))),q(b),q(b),q(f(f(b)))}}
     }
    }
   }
   }
\end{array}
&
\begin{array}{c}  
 \{p(a),p(f(f(b))),q(b),q(b),q(f(f(b)))\}\\
 \\
        \uparrow\\
\\
 \{p(a),\underline{p(f(b))},q(b),\underline{q(f(b))}\}\\
 \\
 	\uparrow\\
	\\
 \{p(a),\underline{p(b),q(f(b))}\}
 \end{array}
\\
\\
\mathrm{Derivation}
&
\mathrm{Rewriting}    	
\end{array}
$$}
\caption{Multiset rewriting over first order atomic formulas as LO proof construction}
\label{LOmsr}
\end{figure*}
\end{example}
From the previous example, we can observe the following properties.
All derivations built using {\em LO rewrite rules} of
Def. \ref{lorewrite} consist of applications of $\para_r$ and $bc$.
Thus, they have no branching (all derivations form a single line).
The combination of a sequence of applications of the
$\para_r$ rule and of the backchaining rule has the following effect:
the head of a ground instance of a rule in $P$ is matched against a sub-multiset
in the current goal; the selected multiset is replaced by the body of the rule.
Clearly, this property allows us to {\em simulate} multiset rewriting over first order
atomic formulas by using LO rewrite rules.

Now, let $F_1$ be the clause 
$$
  p(a)\para p(f(f(b)))\para q(b)\para q(b)\para q(f(f(b)))\lollo \top
$$
If we enrich $P$ with $F_1$, then we can transform the partial derivation of
Figure \ref{LOmsr} into an LO proof as shown below (where $\delta$ stands for
the derivation fragment of Figure \ref{LOmsr}):
$$
   \infer[bc]{\delta}
  {\infer[\top_r]{\dedloPS{\top}}{}}
$$
It is important to note that the same effect can be achieved by adding any formula
with $\top$ that contains a {\em sub-multiset} of the right-hand side of the 
last sequent in the derivation of Figure \ref{LOmsr}.
As an example, let $F_2$ be the formula
$$
  p(a)\para q(b)\lollo \top
$$
If we enrich $P$ with $F_2$, then we can transform the partial derivation of
Figure \ref{LOmsr} into an LO proof as shown below (again, $\delta$ stands for
the derivation fragment of Figure \ref{LOmsr}):
$$
   \infer[bc]{\delta}
  {\infer[\top_r]{\dedloPS{\top,p(f(f(b))),q(b),q(f(f(b)))}}{}}
$$

More in general, let $P$ be a set of LO rewrite rules over $\Sigma$ and
$\calV$, and $\calM,\calM'$ two multisets of ground atomic
formulas (two configurations). Furthermore, let $H$, $G$ the
(possibly empty) $\,\para$-disjunctions of ground atomic formulas
such that $\ms{H}=\calM'$ and $\ms{G}=\calM$. Then, the
provability of the sequent  $\dedLO{P,H\lollo\top}{G}$ precisely
characterizes the problem of {\em coverability} for the  multiset (configuration) 
$\calM'$, namely $\dedLO{P,H\lollo\top}{G}$ is provable if and only if 
there exists a sequence of multiset rewriting steps defined over the theory $P$
that, starting from $\calM$, reaches  a configuration $\calN$ that {\em covers} 
$\calM'$, i.e., such that $\calM'\submult\calN$.

This is a straightforward consequence of the properties of clauses like $H\lollo\top$
(it succeeds only if a sub-multiset of the right-hand side of the current 
sequent matches $\ms{H}$) and of
the fact that, when working with LO rewrite rules, derivations have no branching.
In other words the only way we can transform a partial derivation like the one in Figure
\ref{LOmsr} into a {\em proof} is to apply (once and only once since derivations form a
single line) the clause with $\top$ (i.e., the target configuration is reached).

Coverability is strictly related to  the verification problem of 
{\em safety properties} for concurrent systems \cite{ACJT96,FS01}.
For instance, as shown in \cite{BDM02a}, 
this property allows one to describe properties like {\em coverability} for a marking of 
a Petri Net. In Section \ref{llmcsec}, we will show how to exploit this property in 
the more general case of {\em first order} specifications. 

In Section \ref{reachability} we will discuss a possible characterization of 
{\em reachability} for two configurations using derivability in an extension of LO.

We conclude this section by discussing how {\em universal quantification} can be used
in order to enrich the expressiveness of LO rewrite rules.
\paragraph{The Role of Universal Quantification}
In the {\em proofs as computations} interpretation of logic programs,
universal quantification
is a logical operator which provides a way to generate {\em new values}.
From a logical perspective, this view of universal quantification is based
on its proof-theoretical semantics in intuitionistic 
logic \cite{MNPS91}.
We will define first order rewrite rules with
universal quantification taking inspiration from \cite{CDLMS99}, where
a similar logic fragment, called MSR, is defined. In \cite{CDLMS99},
MSR is used for the specification and analysis of {\em security protocols}.

Given the direct relationship between (first order) multiset rewriting
and (first order) linear logic, it should be evident that multiset rewriting
with universal quantification is the counterpart of LO with universal
quantification.
Having this idea in mind,
we extend the notion of LO rewrite rule as follows.
\begin{definition}\label{qlorewrite}
We call {\em LO quantified rewrite rule} any LO formula having the following form
  $$
  \forall(A_1\para\ldots\para A_n~\lollo \forall x_1,\ldots,x_n.(B_1\para\ldots\para B_m))
  $$
where $A_1,\ldots,A_n$ and $B_1,\ldots,B_m$ are atomic formulas over $\Sigma$ and $\calV$.
\end{definition}
The operational semantics of LO theories consisting of LO quantified rewrite rules
should be clear by looking at the LO proof rule for universally quantified
goal formulas: they are eliminated by introducing new constants.
This operational behavior naturally corresponds to the extension of multiset rewriting with
fresh name generation defined in \cite{CDLMS99}.

\begin{remark}\rm
As mentioned at the beginning of this section, we remark that in \cite{CDLMS99} the logic 
MSR is compared with a fragment of linear logic which turns out to be {\em dual} with respect to
ours, and therefore {\em existential} quantification is used in place of
universal quantification.
Specifically,  an MSR rule is defined as $A \lolli\exists x\ldotp B$, meaning
that $A$ evolves into $B$ by creating a new name for $x$.
In LO with  universal quantification the same effect is obtained via the
clause $A \lollo\forall x\ldotp B$.
In fact, in the goal driven proof system of LO a
computation step is obtained by resolution (i.e., reducing the conclusion
of a clause to its premise).
\end{remark}

The reader may refer to \cite{Cer94b,Cer95,CDLMS99,CDKS00} for a more formal treatment of
the relationship between multiset rewriting and LO.

\subsectionl{Specification of Concurrent Systems}{testlocksec}
The connection with multiset rewriting allows us to think about
LO as a specification language for concurrent systems.
We will illustrate this idea with the help of the following example.
We consider here a distributed {\em test-and-lock} protocol for a net with multiple resources, each of which is
controlled by a monitor.

The protocol is as follows. A set of {\em resources}, distinguished by means
of {\em resource identifiers}, and an arbitrary set of processes are given.
Processes can non-deterministically request access to any resource.
Access to a given resource must be exclusive (only one process at a time).
Mutual exclusion is enforced by providing each resource with a {\em semaphore}.

Given a propositional symbol $init$,
we can encode the initial states of the system as follows:
$$\begin{array}{l}
1\ldotp\ \ init\lollo init\para think\\
[\smallskipamount]
2\ldotp\ \ init\lollo init\para m(x,unlocked)\\
[\smallskipamount]
3\ldotp\ \ init\lollo\anti
\end{array}$$
The atom $think$ represents a thinking (idle) process, while the first order
atom $m(x,s)$ represents a {\em monitor} for the resource with identifier
$x$ and associated semaphore $s$. The semaphore $s$ can assume one of
the two values $locked$ or $unlocked$. Clause 1 and clause 2 can modify the
initial state by adding, respectively, an arbitrary number of thinking
processes and an arbitrary number of resources (with an initially
unlocked semaphore).
Finally, using clause 3 the atom $init$ can be removed after the initialization
phase.

The core of the protocol works as follows:
$$\begin{array}{l}
4\ldotp\ \ think\lollo wait(x)\\
[\smallskipamount]
5\ldotp\ \ wait(x)\lollo think\\
[\smallskipamount]
6\ldotp\ \ wait(x)\para m(x,unlocked)\lollo use(x)\para m(x,locked)\\
[\smallskipamount]
7\ldotp\ \ use(x)\para m(x,locked)\lollo think\para m(x,unlocked)
\end{array}$$
Using clause 4, a process can non-deterministically request access to any
resource with identifier $x$, moving to a waiting state represented by the
atom $wait(x)$. Clause 5 allows a process to go back to thinking from a waiting
state. By clause 6, a waiting process can synchronize with the relevant
monitor and is granted access provided the corresponding semaphore is
unlocked. As a result, the semaphore is locked.
The atom $use(x)$ represents a process which is currently using the resource
with identifier $x$. Clause 7 allows a process to release a resource and
go back to thinking, unlocking the corresponding semaphore.

\begin{remark}\rm
In the previous specification we have intentionally introduced a flaw which we will disclose later
(see Section \ref{examplessec}). Uncovering of this flaw will allow us to explain and better
motivate the use of the universal quantifier for the generation of new names.
\end{remark}

\subsectionl{Linear Logic and Model Checking}{llmcsec}
One of the properties we would like to establish for the specification given
in the previous example is that it ensures mutual exclusion for any resource
used in the system.
One of the difficulties for proving this kind of properties is that the
specification taken into consideration has an infinite number of possible
configurations (all possible rewritings of the goal $init$).

In this paper we will define techniques that can be used to attack this kind of verification
problems by exploiting an interesting connection between verification and
bottom-up evaluation of LO programs.

Let us consider again the protocol specification given in Example
\ref{testlocksec}. The mutual exclusion property can be formulated
as the following property over reachable configurations. Let $S$
be the set of multiset of atomic formulas (i.e., a configuration)
reachable in any derivation from the goal $init$. The protocol
ensures mutual exclusion for resource $x$ if and only if for any
$\calA\in S$, i.e., any reachable configuration,
$\{use(x),use(x)\}$ is not a sub-multiset of $\calA$. In other
words, all goal formulas containing two occurrences of the formula
$use(x)$ represent possible violations of mutual exclusion for
resource $x$.

Following from the previous observation, a possible way of proving
mutual exclusion for our sample protocol is to show that no
configurations having the form $\{use(x),use(x),\ldots\}$ can be
reached starting from the initial states.
This verification methodology can be made effective using a
backward exploration of a protocol specification as described in
\cite{ACJT96}. Specifically, the idea is to saturate the set of
predecessor configurations (i.e., compute all possible predecessor
configurations of the potential violations) and then check that no
initial state occurs in the resulting set.

This verification strategy can be reformulated in a natural way in
our fragment of linear logic. First of all, LO formulas with the
$\all$ constant can be used to finitely represent all possible
violations as follows:
$$\begin{array}{l}
U \eqdef \foralx~{use(x)\para use(x)\lollo\all}
\end{array}$$
Backward reachability amounts then to
compute all possible {\em logical consequences} of the LO
specification of the protocol and of the formula $U$. In logic
programming this strategy is called {\em bottom-up provability}.
If the goal $init$ is in the resulting set, then there exists an
execution (derivation) terminated by an instance of the axiom
$\all$ that leads from $init$ to a multiset of the form
$use(x),use(x),\calA$ for some $x$ and some multiset of atomic
formulas $\calA$.
Thus, the use of clauses with $\all$ to represent violations (and
admissibility of weakening) allows us to reason independently of
the number of processes in the initial states. Following
\cite{ACJT96}, formulas like $\forall x\ldotp use(x)\para use(x)\lollo\all$ can be
viewed as a symbolic representation of {\em upward-closed} sets of
configurations.

On the basis of these observations, the relationship between
reachability and derivability sketched in the previous sections
can be extended as shown in Figure \ref{prerelfig}.
\begin{figure*}[t]
\begin{center}
\begin{tabular}{c|c}
   {\bf Infinite State Concurrent Systems} & {\bf Linear Logic Specification}\\
                     & \\
  transition system  & LO program and proof system\\
  transition         & rule instance\\
  current state      & goal formula\\
  initial state      & initial goal\\
  upward-closed set of states & LO clause with $\top$\\
  forward reachability       & top-down provability\\
  backward reachability      & bottom-up provability
\end{tabular}
\end{center}
\captionl{Reachability versus provability}{prerelfig}
\end{figure*}

In order to exploit this connection and extend the backward
reachability strategy in the rest of the paper we will define a
{\em bottom-up semantics} for first order LO programs. We will
define our semantics via a  fixpoint operator similar to the $T_P$
operator used for logic programs. The fixpoint semantics will give
us an effective way to evaluate bottom-up an LO program, and thus
solve verification problems for infinite-state concurrent systems
as the one described in this section.

\sectionl{A Bottom-up Semantics for \LOf}{fobottomupsec}
The proof-theoretical semantics for \LOf corresponds to the {\em top-down}
operational semantics based on resolution for traditional logic
programming languages like Prolog.
In this paper we are interested in finding a suitable definition of
{\em bottom-up}
semantics that can be used as an alternative operational semantics for \LOf
programs. More precisely, we will define an {\em effective} and
{\em goal-independent} procedure to compute
all goal formulas which are provable from a given program $P$.
This semantics extends the one described in \cite{BDM02a}, which was limited
to propositional LO programs.
In the following, given an \LOf program $P$,
we denote by $\SigmaP$ the signature comprising the set
of constant, function, and predicate symbols in $P$.

\subsectionl{Non-ground Semantics for \LOf}{nongroundsec}
Before discussing the bottom-up semantics, we
lift the definition of operational semantics to \LOf programs.
Following \cite{BDM02a}, we would like to define the operational semantics of a
program $P$ as the set of multisets of {\em atoms} which are provable from $P$.
This could be done by considering the {\em ground instances} of \LOf program
clauses (see Definition \ref{logroundinstdef}).
However, in presence of universal quantification in goals, this solution is
not completely satisfactory. Consider, in fact, the following example.
Take a signature with a
predicate symbol $p$ and two constants $a$ and $b$, and consider the \LOf
program consisting of the axiom $\foralx{p(x)}\lollo\all$
and the program consisting of the two clauses $p(a)\lollo\all$ and
$p(b)\lollo\all$. The two programs would have the same {\em ground} semantics
(i.e., consisting of the two singleton multisets $\{p(a)\}$ and $\{p(b)\}$).
However, the \LOf goal $\foralx{p(x)}$ succeeds only in the
first program, as the reader can verify.
In order to distinguish the two programs, we need to consider the
{\em non ground} semantics. In particular, our aim in this section will be
to extend the so-called {\em C-semantics} of \cite{FLMP93}
to first order LO.

First of all, we give the following definition.
\begin{definition}[Clause Variants]
\label{fovariantdef}
Given an \LOf program $P$,
the set of variants of clauses in $P$,
denoted $\vrn{P}$, is defined as follows:
$$\begin{array}{l}
  \vrn{P}\eqdef\{(H\lollo G)\,\theta\st\forall\,(H\lollo G)
  \in P\ \hbox{and}\ \theta\ \hbox{is a renaming}\\
  \ \ \ \ \ \ \ \ \ \ \ \ \ \ \ \,
  \hbox{of the variables in}\ \fvar{H\lollo G}
  \ \hbox{with new variables}\}\ldotp
\end{array}$$
\end{definition}

Now, we need to reformulate the
proof-theoretical semantics of Section \ref{losec} (see Figure
\ref{system_for_lof}). According to the C-semantics of \cite{FLMP93},
our goal is to define the set of {\em non ground} goals which are provable
from a given program $P$ with an {\em empty answer substitution}.
Slightly departing from \cite{FLMP93}, we modify the proof system of 
Figure \ref{system_for_lof} as follows.
Sequents are defined now over non ground goals. 
The backchaining rule of Figure \ref{system_for_lof} is replaced by the 
new rule shown in Figure \ref{csystem_for_lof} (where, as usual, $\calA$
denotes a multiset of atomic formulas).
The right-introduction rules and the axioms are as in Figure \ref{system_for_lof}.
\begin{figure*}
$$
\begin{array}{c}
\infer[bc\ \ {\mbox {($H\lollo G)\ \in\vrn{P}$}}]
{\dedloPS{\ms{H}\theta,\calA}}
{\dedloPS{G\theta,\calA}}
\end{array}
$$
\caption{Backchaining rule working over non ground goals}
\label{csystem_for_lof}
\end{figure*}
This proof system is based on the
idea of considering a first order program as the (generally {\em infinite})
collection of ({\em non ground})
instances of its clauses. By {\em instance} of a clause
$H\lollo G$, we mean a clause $H\theta\lollo G\theta$, where $\theta$ is
{\em any} substitution.
The reader can see that, with this intuition, the set of goals
provable from the system modified with the backchaining rule shown 
in Figure \ref{csystem_for_lof} corresponds
to the set of non ground goals which are provable with an empty answer
substitution according to \cite{FLMP93}.
This formulation of the proof system is the
proof-theoretical counterpart of the bottom-up semantics
we will define in the following.

All formulas
(and also substitutions) on the right-hand side of the sequents in the proof system
obtained from Figure \ref{system_for_lof} by replacing the 
backchaining rule with the rule of Figure \ref{csystem_for_lof} are implicitly
assumed to range over the set of {\em non ground} terms over $\Sigma$.
Every time rule $\forall_r$ is fired, a new
constant $c$ is added to the current signature, and the resulting goal is
proved in the new signature (see Remark \ref{extrusion}).
Rule $bc$ denotes a backchaining (resolution) step, where $\theta$ indicates
{\em any} substitution.
For our purposes, we can assume $\Dom{\theta}\subseteq\fvar{H}\Union\fvar{G}$
(we remind that $\fvar{F}$ denotes the {\em free} variables of $F$).
Note that $H\lollo G$ is assumed to be a variant, therefore it has
no variables in common with $\calA$.
According to the usual concept of {\em uniformity},
$bc$ can be executed only if the right-hand side of
the current sequent consists of atomic formulas.
Rules $\all_r$, $\,\para_r$, $\with_r$ and $\anti_r$
are the same as in propositional LO.
A sequent is provable if all branches of its proof tree
terminate with instances of the $\all_r$ axiom.

Clearly, the proof system
obtained by considering the rule of Figure \ref{csystem_for_lof} is not {\em effective}, 
however it will be
sufficient for our purposes. An effective way to compute the set of goals which
are provable from the above proof system will be discussed in Section
\ref{foeffectivesec}.

We give the following definition, where $\ded_{\Sigma}$ denotes the provability
relation defined by the proof system of Figure \ref{system_for_lof} in which 
the backchaining rule has been replaced by the rule of Figure \ref{csystem_for_lof}.
\begin{definition}[Operational Semantics]
\label{foopsemdef}
Given an \LOf program $P$, its operational semantics,
denoted $\osP$, is given by
  $$ \osP \eqdef \{\calA\st\calA\ \hbox{is a multiset of (non ground) atoms in}\
              \AsigmaPV\ \hbox{and}\ \dedloP{\SigmaP}{\calA}\}\ldotp $$
\end{definition}
Intuitively, the set $\osP$ is closed by {\em instantiation}, i.e.,
$\calA\theta\in\osP$ for any substitution $\theta$, provided $\calA\in\osP$.
Note that the operational semantics only include multisets of (non ground)
{\em atoms}, therefore no connective (including the {\em universal quantifier})
can appear in the set $\osP$. However, the intuition will be that the variables
appearing in a multiset in $\osP$ must be implicitly considered
{\em universally quantified} (e.g., $\{p(x),q(x)\}\in\osP$ implies that the goal
$\foralx{(p(x)\para q(x))}$ is provable from $P$).
Also note that the information on provable {\em facts} from a given program $P$
is all we need to decide whether a general goal
(possibly with nesting of connectives) is provable from
$P$ or not. In fact, according to \LOf proof-theoretical semantics,
provability of a compound goal can always be
reduced to provability of a finite set of atomic multisets.

\subsectionl{Fixpoint Semantics for \LOfws}{fixpointsec}
We will now discuss the {\em bottom-up} semantics.
In order to deal with universal quantification
(and therefore signature augmentation),
we extend the definitions of Herbrand base and (concrete)
interpretations given in \cite{BDM02a} as follows.
Let $\SigP$ be the set of all possible extensions of the signature $\Sigma_P$
associated to a program $P$ with new constants.
The definition of Herbrand base now depends explicitly on the
signature, and interpretations can be thought of as infinite tuples, with
one element for every signature $\Sigma\in\SigP$.
From here on the powerset of a given set $D$ will be indicated as $\Part{D}$.

We give then the following definitions.%
\begin{definition}[Herbrand Base]
Given an \LOf program $P$ and a signature $\Sigma\in\SigP$, the Herbrand base
of $P$ over $\Sigma$, denoted $\hbS{P}$, is given by
  $$ \hbS{P}\eqdef\MS{\AsigmaV}=\{\calA\st\calA\ \hbox{is a multiset of (non ground)
     atoms in}\ \AsigmaV\}\ldotp $$
\end{definition}
\begin{definition}[Interpretations]
\label{fointerpdef}
Given an \LOf program $P$,
a (concrete) interpretation is a family of sets $\Fam{I}$,
where $I_\Sigma\in\Part{\hbS{P}}$ for every $\Sigma\in\SigP$.
\end{definition}
In the following we often
use the notation $I$ for an interpretation to denote the
family $\Fam{I}$.

Interpretations form a complete lattice where inclusion and least upper bound
are defined like (component-wise) set inclusion and union. In the following
definition we therefore overload the symbols $\subseteq$ and $\Union$ for sets.
\begin{definition}[Interpretation Domain]
\label{fodomaindef}
Interpretations form a complete lattice
$\couple{\calD}{\subseteq}$, where:
\begin{itemize}
  \item $\calD=\{I\st I\ \hbox{is an interpretation}\}$;
  \item $I\subseteq J$ if and only if $I_\Sigma\subseteq J_\Sigma$ for every
    $\Sigma\in\SigP$;
  \item the least upper bound of $I$ and $J$ is
    $\{I_\Sigma\Union J_\Sigma\}_{\Sigma\in\SigP}$;
  \item the bottom and top elements are
    $\eset=\Fam{\eset}$ and
    $\{\hbS{P}\}_{\Sigma\in\SigP}$, respectively.
\end{itemize}
\end{definition}
Before introducing the definition of fixpoint operator,
we need to define the notion of satisfiability of a context $\Delta$ (a multiset of 
goal formulas) in a
given interpretation $I$. For this purpose, we introduce the judgment
$\valS{I}{\Delta}{\calC}$,
where $I$ is an {\em input} interpretation, $\Delta$ is an {\em input} context,
and $\calC$ is an {\em output}
fact (a multiset of atomic formulas). 
The judgment is also parametric with respect to a given signature $\Sigma$.

The need for this judgment, with respect to the familiar logic programming
setting \cite{GDL95}, is motivated by the arbitrary nesting of
connectives in \LOf clause bodies. The satisfiability judgment is modeled
according to the right-introduction rules of the connectives. In other words,
the computation performed by the satisfiability judgment corresponds to
{\em top-down} steps inside our {\em bottom-up} semantics. Intuitively,
the parameter $\calC$ must be thought of as
an {\em output} fact such that $\calC+\Delta$ is valid in $I$.
The notion of output fact will simplify the
presentation of the algorithmic version of the judgment which we will present
in Section \ref{foeffectivesec}.
The notion of {\em satisfiability} is modeled according to the
right-introduction (decomposition) rules of the proof system, as follows
(we remind that '+' denotes multiset union).
\begin{definition}[Satisfiability Judgment]
\label{satjdef}
Let $P$ be an \LOf program,
$\Sigma\in\SigP$, and $I=\Fam{I}$ an interpretation.
The satisfiability judgment $\satS$ is defined as follows:
$$\begin{array}{l}
  \valS{I}{\all,\Delta}{\calC}\ \hbox{for any fact}\ \calC\ \hbox{in}\
  \AsigmaV;\\
  [\medskipamount]
  \valS{I}\calA{\calC}\ \hbox{if}\ \calA+\calC \in I_\Sigma;\\
  [\medskipamount]
  \valS{I}{\foralx{G},\Delta}{\calC}\ \hbox{if}\
  \val{I}{\Sigma,c}{G[c/x],\Delta}{\calC},\ \hbox{with}\ c\not\in\Sigma
  \ \hbox{{\rm (}see remark \ref{satexportremark}{\rm )}};\\
  [\medskipamount]
  \valS{I}{G_1\with G_2,\Delta}{\calC}\ \hbox{if}\
  \valS{I}{G_1,\Delta}{\calC}\ \hbox{and}\
  \valS{I}{G_2,\Delta}{\calC};\\
  [\medskipamount]
  \valS{I}{G_1\para G_2,\Delta}{\calC}\ \hbox{if}\
  \valS{I}{G_1,G_2,\Delta}{\calC};\\
  [\medskipamount]
  \valS{I}{\anti,\Delta}{\calC}\ \hbox{if}\
  \valS{I}{\Delta}{\calC}\ldotp
\end{array}$$
\end{definition}
\begin{remark}\rm
\label{satexportremark}
When using the notation $\valS{I}{\Delta}{\calC}$
we {\em always} make the {\em implicit} assumption that $\Delta$ is
a context defined over $\Sigma$ (i.e., term constructors in $\Delta$ must
belong to $\Sigma$). As a result, also the output fact $\calC$ must be
defined over $\Sigma$. This assumption, which is the counterpart
(see Remark \ref{extrusion}) of an
analogous assumption for proof systems like the one in Figure
\ref{system_for_lof}, i.e., with explicit signature notation,
will {\em always} and {\em tacitly} hold in the following.
For example, note that in the $\forall$-case of the $\satS$ definition below,
the newly introduced constant $c$ {\em cannot be exported} through the
output fact $\calC$. This is crucial to capture the operational semantics
of the universal quantifier.
\end{remark}

The satisfiability judgment $\satS$ satisfies the following properties.
\begin{lemma}
\label{fomovelemma}
For every interpretation $I=\Fam{I}$,
context $\Delta$, and fact $\calC$,
  $$ \valS{I}{\Delta}{\calC}\ \hbox{if and only if}\ \valS{I}{\Delta,\calC}{\Eps}\ldotp $$
\end{lemma}
\begin{proof}
See Appendix \ref{proofsapp}.
\end{proof}
\begin{lemma}
\label{fovallemma}
For any interpretations $I_1=\Fam{(I_1)}$, $I_2=\Fam{(I_2)}$, \ldots,
context $\Delta$, and fact $\calC$,
\begin{enumerate}
  \item if $I_1\subseteq I_2$ and $\valS{I_1}{\Delta}{\calC}$ then
    $\valS{I_2}{\Delta}{\calC}$;
  \item if $I_1\subseteq I_2\subseteq\ldots$ and
    $\valS{\Un{i}{\Inf}I_i}{\Delta}{\calC}$ then there exists $k\in\Nat$
    s.t. $\valS{I_k}{\Delta}{\calC}$.
\end{enumerate}
\end{lemma}
\begin{proof}
See Appendix \ref{proofsapp}.
\end{proof}
We are now ready to define the fixpoint operator $\Tp$.
\begin{definition}[Fixpoint Operator {\boldmath $\Tp$}]
\label{Tpdef}
Given an \LOf program $P$ and an interpretation $I=\Fam{I}$,
the fixpoint operator $\Tp$ is defined as follows:
$$
\begin{array}{l}
  \Tp(I)\eqdef\Fam{(\Tp(I))};\\
  [\medskipamount]
  (\Tp(I))_\Sigma\eqdef\{\ms{H}\theta+\calC\st
  (H\lollo G)\in\vrn{P},\ \theta\ \hbox{is}\\
  \ \ \ \ \ \ \ \ \ \ \ \ \ \ \ \
  \hbox{any substitution,}\ \hbox{and}\
  \valS{I}{G\theta}{\calC}\}\ldotp\\
\end{array}
$$
\end{definition}
\begin{remark}\rm
In the previous definition, $\theta$ is implicitly assumed to be defined over
$\Sigma$, i.e., $\theta$ can only map variables in $\Dom{\theta}$ to terms in
$\TsigmaX{\calV}$.
\end{remark}
The following property holds.
\begin{proposition}[Monotonicity and Continuity]
\label{monconttpprop}
For every \LOf program $P$, the fixpoint operator $\Tp$ is monotonic and
continuous over the lattice $\couple{\calD}{\subseteq}$.
\end{proposition}
\begin{proof}
{\em Monotonicity}.\nl
Immediate from the definition of $\Tp$ and item $i$ of Lemma \ref{fovallemma}.
\vsep
{\em Continuity}.\nl
We prove that
$\Tp$ is finitary, i.e., for any sequence of interpretations
$I_1\subseteq I_2\subseteq\ldots$ we have that
$\Tp(\Un{i}{\Inf}I_i)\subseteq\Un{i}{\Inf}\Tp(I_i)$, i.e.,
for every $\Sigma\in\SigmaP$,
$(\Tp(\Un{i}{\Inf}I_i))_\Sigma\subseteq(\Un{i}{\Inf}\Tp(I_i))_\Sigma$.
Let $\calA\in(\Tp(\Un{i}{\Inf}I_i))_\Sigma$.
By definition of $\Tp$, there exist a variant
$H\lollo G$ of a clause in $P$, a substitution $\theta$, and
a fact $\calC$ s.t.
$\calA=\ms{H}\theta+\calC$ and $\valS{\Un{i}{\Inf}I_i}{G\theta}{\calC}$.
By item $ii$ of Lemma \ref{fovallemma}, we have that there exists
$k\in\Nat$ s.t. $\valS{I_k}{G\theta}{\calC}$. Again by definition of $\Tp$,
we get $\calA=\ms{H}\theta+\calC\in(\Tp(I_k))_\Sigma\subseteq
(\Un{i}{\Inf}{\Tp(I_i)})_\Sigma$.
\end{proof}
Monotonicity and continuity of the $\Tp$ operator imply, by Tarski's Theorem,
that $\lfp{\Tp}=\itp{\omega}$. The fixpoint semantics of a program $P$ is
then defined as follows.
\begin{definition}[Fixpoint Semantics]
Given an \LOf
program $P$, its fixpoint semantics, denoted $\fsP$, is defined as
follows:
  $$ \fsP\eqdef(\lfp{\Tp})_\SigmaP=(\Itp{\omega}{\Fam{\eset}})_\SigmaP\ldotp $$
\end{definition}
We conclude this section by proving the
following fundamental result, which states that the fixpoint semantics is sound
and complete with respect to the operational semantics (see Definition \ref{foopsemdef}).
\begin{theorem}[Soundness and Completeness]
\label{fofixpoint_operational}
For every \LOf program $P$, $\fsP$ $=$ $\osP$.
\end{theorem}
\begin{proof}
$\fsP\subseteq\osP$.
\nl
We prove that for every $k\in\Nat$, for every signature
$\Sigma\in\SigP$, and for every context $\Delta$,
$\valS{\itp{k}}{\Delta}{\Eps}$ implies
$\dedloPS{\Delta}$. The proof is by lexicographic induction on $(k,h)$, where
$h$ is the length of the derivation of $\valS{\itp{k}}{\Delta}{\Eps}$.
\begin{itemize}
  \item[-] If $\Delta=\all,\Delta'$, obvious;
  \item[-] if $\Delta=\calA$ and $\calA\in(\itp{k})_\Sigma$, then there exist
    a variant $H\lollo G$ of a clause in $P$, a fact $\calC$
    and a substitution $\theta$ s.t.
    $\calA=\ms{H}\theta+\calC$ and $\valS{\itp{k-1}}{G\theta}{\calC}$.
    By Lemma \ref{fomovelemma}, this implies
    $\valS{\itp{k-1}}{G\theta,\calC}{\Eps}$.
    Then by the inductive hypothesis we have $\dedloPS{G\theta,\calC}$, from
    which $\dedloPS{\ms{H}\theta,\calC}$, i.e., $\dedloPS{\calA}$ follows
    by $bc$ rule;
  \item[-] if $\Delta=\foralx{G},\Delta'$ and
    $\val{\itp{k}}{\Sigma,c}{G[c/x],\Delta'}{\Eps}$, with $c\not\in\Sigma$,
    then by the inductive hypothesis we have $\dedloP{\Sigma,c}{G[c/x],\Delta'}$
    from which $\dedloPS{\foralx{G},\Delta'}$ follows by $\forall_r$ rule;
  \item[-] if $\Delta=G_1\with G_2,\Delta'$,
    $\valS{\itp{k}}{G_1,\Delta'}{\Eps}$, and
    $\valS{\itp{k}}{G_2,\Delta'}{\Eps}$, then
    by the inductive hypothesis we have
    $\dedloPS{G_1,\Delta'}$ and $\dedloPS{G_2,\Delta'}$, from which
    $\dedloPS{G_1\with G_2,\Delta'}$ follows by $\with_r$ rule;
  \item[-] if $\Delta=G_1\para G_2,\Delta'$ and
    $\valS{\itp{k}}{G_1,G_2,\Delta'}{\Eps}$, then
    by the inductive hypothesis we have $\dedloPS{G_1,G_2,\Delta'}$, from
    which $\dedloPS{G_1\para G_2,\Delta'}$ follows by $\para_r$ rule;
  \item[-] if $\Delta=\anti,\Delta'$ and
    $\valS{\itp{k}}{\Delta'}{\Eps}$, then
    by the inductive hypothesis we have $\dedloPS{\Delta'}$, from
    which $\dedloPS{\anti,\Delta'}$ follows by $\anti_r$ rule.
\end{itemize}
$\osP\subseteq\fsP$.
\nl
We prove that for every signature $\Sigma\in\SigP$ and for every context
$\Delta$, if $\dedloPS{\Delta}$ then there exists $k\in\Nat$ s.t.
$\valS{\itp{k}}{\Delta}{\Eps}$. The proof is by induction on the derivation of
$\dedloPS{\Delta}$.
\begin{itemize}
  \item[-] If $\Delta=\all,\Delta'$, then for every $k\in\Nat$,
    $\valS{\itp{k}}{\Delta}{\Eps}$;
  \item[-] if $\Delta=\ms{H}\theta,\calA$, with $H\lollo G$ a variant of a clause
    in $P$, $\theta$ substitution, and $\dedloPS{G\theta,\calA}$, then
    by the inductive hypothesis we have that there exists $k\in\Nat$ s.t.
    $\valS{\itp{k}}{G\theta,\calA}{\Eps}$. Then, by Lemma \ref{fomovelemma},
    $\valS{\itp{k}}{G\theta}{\calA}$. By definition of $\Tp$,
    $\ms{H}\theta+\calA\in(\itp{k+1})_\Sigma$, which implies
    $\valS{\itp{k+1}}{\ms{H}\theta+\calA}{\Eps}$;
  \item[-] if $\Delta=\foralx{G},\Delta'$ and
    $\dedloP{\Sigma,c}{G[c/x],\Delta'}$, with $c\not\in\Sigma$, then by
    the inductive hypothesis we have that there exist $k\in\Nat$ s.t.
    $\val{\itp{k}}{\Sigma,c}{G[c/x],\Delta'}{\Eps}$, from which
    $\valS{\itp{k}}{\foralx{G},\Delta'}{\Eps}$ follows;
  \item[-] if $\Delta=G_1\with G_2,\Delta'$,
    $\dedloPS{G_1,\Delta'}$ and $\dedloPS{G_2,\Delta'}$, then
    by the inductive hypothesis we have that there exist $k_1,k_2\in\Nat$ s.t.
    $\valS{\itp{k_1}}{G_1,\Delta'}{\Eps}$ and
    $\valS{\itp{k_2}}{G_2,\Delta'}{\Eps}$. By taking $k=max\{k_1,k_2\}$,
    by item $i$ of Lemma \ref{fovallemma} and monotonicity of $\Tp$ (Proposition \ref{monconttpprop})
    we get
    $\valS{\itp{k}}{G_1,\Delta'}{\Eps}$ and
    $\valS{\itp{k}}{G_2,\Delta'}{\Eps}$, from which
    $\valS{\itp{k}}{G_1\with G_2,\Delta'}{\Eps}$ follows;
  \item[-] if $\Delta=G_1\para G_2,\Delta'$ or $\Delta=\anti,\Delta'$,
    the conclusion follows by a
    straightforward application of the inductive hypothesis.
\end{itemize}
\end{proof}
\begin{example}
Let $\Sigma$ be a signature including the constant symbols $a$ and $b$,
a function symbol $f$, and the predicate symbols $p,q,r$, let $\calV$ be a
denumerable set of variables and $x,y\in\calV$, and let $P$ be the
following \LOf program:
$$\begin{array}{l}
  1\ldotp\ \ r(f(b))\para p(a)\lollo\all\\
  [\smallskipamount]
  2\ldotp\ \ p(x)\lollo\all\\
  [\smallskipamount]
  3\ldotp\ \ q(y)\lollo(\foralx{p(x)})\with r(y)
\end{array}$$
Let $I_0=\Fam{\eset}$, and let us compute $I_1=\Tp(I_0)$.
Using clauses 1 and 2, we get that (see Definitions \ref{satjdef} and \ref{Tpdef})
$(I_1)_\Sigma$ contains the multisets of
atoms of the form $\{r(f(b)),p(a)\}+\calA$, and
$\{p(t)\}+\calA$, where $\calA$ is
any multiset of (possibly non-ground) atoms in $\AsigmaV$, while $t$ is any
(possibly non ground) term in $\TsigmaV$.
Similarly $(I_1)_{\Sigma'}$, for a
generic signature $\Sigma'$ such that $\Sigma\subseteq\Sigma'$, contains all
multisets of the above form where $\calA$ and $t$ are taken from,
respectively, $\AsigV{\Sigma'}$ and $\TsigV{\Sigma'}$. For instance, let
$c$ be a new constant not appearing in $\Sigma$. The set $(I_1)_{\Sigma'}$ will
contain, e.g., the multisets $\{p(c)\}$, $\{p(f(c)),q(b)\}$, and so on.

Now, consider the substitution $\theta=[\substbind{y}{f(b)}]$ and the
following corresponding instance of clause 3:
$q(f(b))\lollo(\foralx{p(x)})\with r(f(b))$.
Assume we want to compute an output fact $\calC$ for the judgment
 $$ \valS{I_1}{(\foralx{p(x)})\with r(f(b))}{\calC}\ldotp $$
By definition of $\satP$, we have to compute
$\valS{I_1}{(\foralx{p(x)})}{\calC}$ and
$\valS{I_1}{r(f(b))}{\calC}$. For the latter judgment we have that, e.g.,
$\valS{I_1}{r(f(b))}{p(a)}$. For the first judgment, by definition of $\satP$,
we must compute $\val{I_1}{\Sigma,c}{{p(c)}}{\calC}$, where $c$ is a
new constant not in $\Sigma$. As $\{p(c)\}$
is contained in $(I_1)_{\Sigma,c}$, we
can get that $\val{I_1}{\Sigma,c}{{p(c)}}{\Eps}$. We can also get
$\val{I_1}{\Sigma,c}{{p(c)}}{p(a)}$ (in fact $\{p(c),p(a)\}$ is also
contained in $(I_1)_{\Sigma,c}$. By applying the $\with$-rule for $\satP$,
we get that $\valS{I_1}{(\foralx{p(x)})\with r(f(b))}{p(a)}$.
Therefore, by applying clause 3 we get that, e.g., the multiset $\{q(b),p(a)\}$ is in
$(I_2)_\Sigma=(\Tp(I_1))_\Sigma$.
\end{example}
\sectionl{An Effective Semantics for \LOf}{foeffectivesec}
The fixpoint operator $\Tp$ defined in the previous section
does not enjoy one of the crucial properties we required for our
bottom-up semantics, namely its definition is {\em not} effective.
This is a result of both the definition of the satisfiability judgment
(whose clause for $\all$ is clearly not effective) and the definition of
interpretations as infinite tuples. In order to solve these problems, we first
define the (abstract) Herbrand base and (abstract) interpretations as
follows.
\begin{definition}[Abstract Herbrand Base]
Given an \LOf
program $P$, the Herbrand
base of $P$, denoted $\hb{}{P}$, is given by
  $$ \hb{}{P}\eqdef\hb{\SigmaP}{P}\ldotp $$
\end{definition}
\begin{definition}[Abstract Interpretations]
\label{foainterpdef}
Given an \LOf  program $P$, an
interpretation $I$ is any subset of $\hb{}{P}$,
i.e., $I\in\Part{\hb{}{P}}$.
\end{definition}
In order to define the abstract domain of interpretations, we need the
following definitions.
\begin{definition}[Instance Operator]
\label{foinstopdef}
Given an interpretation $I$ and a signature $\Sigma\in\SigP$,
we define the operator $\instS$ as follows:
  $$ \InstS{I}=\{\calA\theta\st\calA\in I,\ \theta\ \hbox{substitution over}\
     \Sigma\}\ldotp $$
\end{definition}
\begin{definition}[Upward-closure Operator]
\label{foupclosdef}
Given an interpretation $I$ and a signature $\Sigma\in\SigP$,
we define the operator $\upS$ as follows:
  $$ \UpS{I}=\{\calA+\calC\st\calA\in I,\ \calC\ \hbox{fact over}\ \Sigma\}\ldotp $$
\end{definition}
\begin{remark}\rm
Note that, as usual,
in the previous definitions we assume the substitution
$\theta$ and the fact $\calC$ to be defined over the signature $\Sigma$.
\end{remark}
The following definition provides the connection between the (abstract)
interpretations defined in Definition \ref{foainterpdef} and the (concrete)
interpretations of Definition \ref{fointerpdef}. The idea behind the
definition is that an interpretation implicitly {\em denotes}
the set of elements
which can be obtained by either {\em instantiating}
or {\em closing upwards} elements in the
interpretation itself (where the concepts of instantiation and upward-closure
are made precise by the above definitions). The operation of instantiation
is related to the notion of C-semantics \cite{FLMP93} (see Definition
\ref{foopsemdef}), while the operation of upward-closure is justified by
Proposition \ref{foaffine}. Note that the operations of
instantiation and upward-closure are performed for every possible
signature $\Sigma\in\SigP$.
\begin{definition}[Denotation of an Interpretation]
\label{fodenotdef}
Given an (abstract) interpretation $I$, its denotation $\den{I}$
is the (concrete) interpretation $\Fam{\den{I}}$ defined as follows:
  $$ \den{I}_\Sigma\eqdef\InstS{\UpS{I}}\ \ \
     (\hbox{or,\ equivalently,}\ \den{I}_\Sigma\eqdef\UpS{\InstS{I}}). $$
Two interpretations $I$ and $J$ are said to be equivalent, written
$I\iequiv J$, if and only if $\den{I}=\den{J}$.
\end{definition}
The equivalence of the two different equations in Definition \ref{fodenotdef}
is stated in the following proposition.
\begin{proposition}
For every interpretation $I$, and signature $\Sigma\in\SigP$,
  $$ \InstS{\UpS{I}}=\UpS{\InstS{I}}. $$
\end{proposition}
\begin{proof}
Let $(\calA+\calC)\theta\in\InstS{\UpS{I}}$, with $\calA\in I$. Then
$(\calA+\calC)\theta=(\calA\theta)+\calC\theta\in\UpS{\InstS{I}}$.
Conversely, let $\calA\theta+\calC\in\UpS{\InstS{I}}$, with $\calA\in I$.
Let $\calB$ be a variant of $\calC$ with new variables (not appearing in
$\calA$, $\theta$, and $\calC$) and $\theta'$ be the substitution with
domain $\Dom{\theta}\Union\fvar{\calB}$ and s.t.
$\restr{\theta'}{\Dom{\theta}}=\theta$ and
$\theta'$ maps $\calB$ to $\calC$. Then
$\calA\theta+\calC=\calA\theta'+\calB\theta'=(\calA+\calB)\theta'\in
\InstS{\UpS{I}}$.
\end{proof}
We are now ready to define the symbolic interpretation domain. In the following 
we will use the word {\em  abstract} to stress the connection between our symbolic 
semantics and the theory of {\em abstract interpretation}.
Our abstraction does not loose precision but it allows us to finitely represent 
infinite collections of formulas.
As previously mentioned, the idea is that of considering interpretations as implicitly
defining the sets of elements contained in their denotations. Therefore,
differently from Definition \ref{fodomaindef}, now we need to check containment between
{\em denotations}. Furthermore,
as we do not need to distinguish between interpretations having the same
denotation, we simply identify them using equivalence classes with respect
to the corresponding equivalence relation $\iequiv$.
\begin{definition}[Abstract Interpretation Domain]
\label{foadomaindef}
Abstract interpretations form
a complete lattice $\couple{\calI}{\asubseteq}$, where
\begin{itemize}
  \item $\calI=\{\iclass{I}\st I\ \hbox{is an interpretation}\}$;
  \item $\iclass{I}\asubseteq\iclass{J}$ if and only if $\den{I}\subseteq\den{J}$;
  \item the least upper bound of $\iclass{I}$ and $\iclass{J}$,
    written $\alub{\iclass{I}}{\iclass{J}}$, is $\iclass{I\Union J}$;
  \item the bottom and top elements are
    $\iclass{\eset}$ and $\iclass{\Eps}$, respectively.
\end{itemize}
\end{definition}
The following proposition provides an {\em effective} and  equivalent condition
for testing the $\asubseteq$ relation (which we call {\em entailment} relation)
over interpretations. We will need this result later on.
\begin{proposition}[Entailment between Interpretations]
\label{foasubseteqprop}
Given two interpretations $I$ and $J$, $\den{I}\subseteq\den{J}$ if and only if
for every $\calA\in I$, there exist $\calB\in J$, a substitution
$\theta$, and a fact $\calC$ (defined over $\SigmaP$) s.t.
$\calA=\calB\theta+\calC$.
\end{proposition}
\begin{proof}
{\em If part}. We prove that for every $\Sigma\in\SigP$,
$\den{I}_\Sigma\subseteq\den{J}_\Sigma$.
Let $\calA'=\calA\theta'+\calC'\in\UpS{\InstS{I}}=\den{I}_\Sigma$,
with $\calA\in I$ and
$\theta'$, $\calC'$ defined over $\Sigma$.
By hypothesis, there exist $\calB\in J$, a substitution
$\theta$, and a fact $\calC$ (defined over $\SigmaP$) s.t.
$\calA=\calB\theta+\calC$.
Therefore, $\calA'=\calA\theta'+\calC'=(\calB\theta+\calC)\theta'+\calC'=
\calB\theta\theta'+(\calC\theta'+\calC')\in\UpS{\InstS{J}}=\den{J}_\Sigma$
(note that $\theta\theta'$ and $\calC\theta'+\calC'$ are both defined over
$\Sigma$ because $\SigmaP\subseteq\Sigma$).
\medskip\\
{\em Only if part}.
Let $\calA\in I$, then $\calA\in\den{I}_\SigmaP$ (note that
$\calA$ is defined over $\SigmaP$ by definition of interpretation).
Then, by the hypothesis we have that
$\calA\in\den{J}_\SigmaP=\Up{\SigmaP}{\Inst{\SigmaP}{J}}$, i.e.,
there exist $\calB\in J$, a substitution $\theta$, and a fact $\calC$
(defined over $\SigmaP$) s.t. $\calA=\calB\theta+\calC$.
\end{proof}
We now define the abstract
satisfiability judgment $\avalS{I}{\Delta}{\calC}{\theta}$, where
$I$ is an {\em input} interpretation, $\Delta$ is an {\em input} context, $\calC$ is an
{\em output} fact, and $\theta$ is an {\em output} substitution.
\begin{remark}\rm
\label{asatexportremark}
As usual, the notation $\avalS{I}{\Delta}{\calC}{\theta}$ requires that
$\Delta$, $\calC$, and $\theta$ are defined over the signature $\Sigma$. As a consequence,
the newly introduced constant $c$ in the $\forall$-case of the $\asatS$
definition below {\em cannot be exported} through the output parameters
$\calC$ or $\theta$.
\end{remark}
The judgment $\asatS$ can be thought of as an abstract version of the
judgment $\satS$ (compare Definition \ref{satjdef}).
We now need one more parameter, namely an {\em output}
substitution. The idea behind the definition is that the output fact
$\calC$ and the output substitution $\theta$ are {\em minimal}
(in a sense to be clarified) so that they can be computed effectively given
a program $P$, an interpretation $I$, and a signature $\Sigma$. The output
substitution $\theta$ is needed in order
to deal with clause instantiation, and its
minimality is ensured by using most general unifiers in the definition.
As the reader can note, the sources of non-effectiveness which are present
in Definition \ref{satjdef} (e.g., in the rule for $\all$))
are removed in Definition \ref{foasatdef} below.
We recall that the notation $\theta_1\slub\theta_2$ denotes the least upper
bound of substitutions (see Appendix \ref{prelimsec}).
\begin{definition}[Abstract Satisfiability Judgment]
\label{foasatdef}
Let $P$ be an \LOf program, $I$ an interpretation, and
$\Sigma\in\SigP$. The abstract
satisfiability judgment $\asatS$ is defined as follows:
$$\begin{array}{l}
  \avalS{I}{\all,\Delta}{\Eps}{nil};\\
  [\medskipamount]
  \avalS{I}{\calA}{\calC}{\theta}\ \hbox{if there exist}\ \calB\in I\
  \hbox{(variant)},\ \calB'\submult\calB,\ \calA'\submult\calA,\
  \card{\calB'}=\card{\calA'},\\
  \ \ \ \ \ \ \ \ \ \ \ \ \ \ \ \ \ \ \ \calC=\calB\diff\calB',\ \hbox{and}\ \
  \theta=\restr{\Mmgu{\calB'}{\calA'}}{\fvar{\calA,\calC}};\\
  [\medskipamount]
  \avalS{I}{\foralx{G},\Delta}{\calC}{\theta}\ \hbox{if}\
  \aval{I}{\Sigma,c}{G[c/x],\Delta}{\calC}{\theta},
  \ \hbox{with}\ c\not\in\Sigma
  \ \hbox{{\rm (}see Remark \ref{asatexportremark}{\rm )}};\\
  [\medskipamount]
  \avalS{I}{G_1\with G_2,\Delta}{\calC}{\theta}\ \hbox{if}\
  \avalS{I}{G_1,\Delta}{\calC_1}{\theta_1},\
  \avalS{I}{G_2,\Delta}{\calC_2}{\theta_2},\\
  \ \ \ \ \ \ \ \ \ \ \ \ \ \ \ \ \ \ \ \ \ \ \ \ \ \ \ \ \
  \calD_1\submult\calC_1,\ \calD_2\submult\calC_2,\
  \card{\calD_1}=\card{\calD_2},\ \theta_3=\Mmgu{\calD_1}{\calD_2},\\
  \ \ \ \ \ \ \ \ \ \ \ \ \ \ \ \ \ \ \ \ \ \ \ \ \ \ \ \ \
  \calC=\calC_1+(\calC_2\diff\calD_2),\ \hbox{and}\ \
  \theta=\restr{(\theta_1\slub\theta_2\slub\theta_3)}
               {\fvar{G_1,G_2,\Delta,\calC}};\\
  [\medskipamount]
  \avalS{I}{G_1\para G_2,\Delta}{\calC}{\theta}\ \hbox{if}\
  \avalS{I}{G_1,G_2,\Delta}{\calC}{\theta};\\
  [\medskipamount]
  \avalS{I}{\anti,\Delta}{\calC}{\theta}\ \hbox{if}\
  \avalS{I}{\Delta}{\calC}{\theta}\ldotp
\end{array}$$
\end{definition}
We recall that two multisets in general may have more than one
(not necessarily equivalent) most general unifier
and that using the notation $\Mmgu{\calB'}{\calA'}$ we mean any
unifier which is {\em non-deterministically} picked from the set of most
general unifiers of $\calB'$ and $\calA'$ (see Appendix
\ref{prelimsec}).
\begin{example}
\label{foasatex}
Let us consider a signature with a function symbol $f$ and predicate symbols
$p,q,r,s$. Let $\calV$ be a denumerable set of variables, and
$u,v,w,\ldots\in\calV$.
Let $I$ be the interpretation consisting of the two multisets
$\{p(x),q(x)\}$ and $\{r(y),p(f(y))\}$
(for simplicity, hereafter we omit brackets in
multiset notation), and $P$ the program
$$\begin{array}{l}
  1\ldotp\ \ r(w)\lollo q(f(w))\\
  [\smallskipamount]
  2\ldotp\ \ s(z)\lollo\foralx{p(f(x))}\\
  [\smallskipamount]
  3\ldotp\ \ \anti\lollo q(u)\with r(v)\\
\end{array}$$
Let us consider (a renaming of) the body of the first clause, $q(f(w'))$,
and (a renaming of) the first element in $I$, $p(x'),q(x')$.
Using the second case for the
$\asat{\SigmaP}$ judgment, with $\calA=\calA'=q(f(w'))$,
$\calB=p(x'),q(x')$, $\calB'=q(x')$, we get
  $$ \aval{I}{\SigmaP}{q(f(w'))}{p(x')}{[\substbind{x'}{f(w')}]}\ldotp $$
Let us consider now (a renaming of)
the body of the second case, $\foralx{p(f(x))}$,
and another renaming of the first element, $p(x''),q(x'')$.
From the $\forall$-case of the definition of $\asat{\SigmaP}$,
$\aval{I}{\SigmaP}{\foralx{p(f(x))}}{\calC}{\theta}$ if
$\aval{I}{\SigmaP,c}{p(f(c))}{\calC}{\theta}$, with $c\not\in\SigmaP$.
Now, we can apply the second case for $\asat{\SigmaP,c}$. Unfortunately,
we can't choose $\calA'$ to be $p(f(c))$ and $\calB'$ to be $p(x'')$. In fact,
by unifying $p(f(c))$ with $p(x'')$, we should get the substitution
$\theta=[\substbind{x''}{f(c)}]$
and the output fact $q(x'')$ (note that $x''$ is a free
variable in the output fact)
and this is not allowed because the substitution $\theta$ must be defined on
$\SigmaP$, in order for $\aval{I}{\SigmaP}{\foralx{p(f(x))}}{\calC}{\theta}$
to be meaningful. It turns out that the only way to use the second clause for
$\asat{\SigmaP,c}$ is to choose $\calA'=\calB'=\Eps$, which is useless in
the fixpoint computation (see Example \ref{foSPex}).
Finally, let us consider (a renaming of)
the body of the third clause, $\anti\lollo q(u')\with r(v')$.
According to the $\with$-rule for the $\asat{\SigmaP}$ judgment, we must
first compute $\calC_1$, $\calC_2$, $\theta_1$ and $\theta_2$ such that
$\aval{I}{\SigmaP}{q(u')}{\calC_1}{\theta_1}$ and
$\aval{I}{\SigmaP}{r(v')}{\calC_2}{\theta_2}$.
To this aim, take two variants of the multisets in $I$,
$p(x'''),q(x''')$ and $r(y'),p(f(y'))$.
Proceeding as above, we get that
  $$ \aval{I}{\SigmaP}{q(u')}{p(x''')}{[\substbind{u'}{x'''}]} \ \
     \hbox{and}\ \ \aval{I}{\SigmaP}{r(v')}{p(f(y'))}{[\substbind{v'}{y'}]}\ldotp $$
Now, we can apply the $\with$-rule for the $\asat{\SigmaP}$ judgment,
with $\calD_1=p(x''')$, $\calD_2=p(f(y'))$, and
$\theta_3=[\substbind{x'''}{f(y')}]$.
We have that $\theta_1\slub\theta_2\slub\theta_3=[\substbind{u'}{f(y')},
\substbind{v'}{y'},\substbind{x'''}{f(y')}]$.
Therefore, we get that
  $$ \aval{I}{\SigmaP}{q(u')\with r(v')}{p(x''')}{[\substbind{u'}{f(y')},
     \substbind{v'}{y'},\substbind{x'''}{f(y')}]}\ldotp $$
\end{example}
The following lemma states a simple property of the substitution domain,
which we will need in the following.
\begin{lemma}
\label{fodomainlemma}
For every interpretation $I$, context $\Delta$, fact $\calC$, and substitution
$\theta$, if $\avalS{I}{\Delta}{\calC}{\theta}$ then
$\Dom{\theta}\subseteq\fvar{\Delta}\Union\fvar{\calC}$.
\end{lemma}
\begin{proof}
Immediate by induction on the definition of $\asatS$.
\end{proof}
The connection between the satisfiability judgments $\satS$ and $\asatS$ is
clarified by the following lemma (in the following we denote
by $\supmult$ the converse of the sub-multiset relation, i.e.,
$\calA\supmult\calB$ if and only if $\calB\submult\calA$).
\begin{lemma}
\label{fovalavalrellemma}
For every interpretation $I$, context $\Delta$, fact $\calC$, and substitution
$\theta$,
\begin{enumerate}
  \item if $\avalS{I}{\Delta}{\calC}{\theta}$ then
    $\valS{\den{I}}{\Delta\theta\theta'}{\calC'\theta'}$ for every substitution
    $\theta'$ and fact $\calC'\supmult\calC\theta$;
  \item if $\valS{\den{I}}{\Delta\theta}{\calC}$ then there exist a fact
    $\calC'$, and substitutions $\theta'$ and $\sigma$ s.t.
    $\avalS{I}{\Delta}{\calC'}{\theta'}$,
    $\restr{\theta}{\fvar{\Delta}}=
    \restr{(\compos{\theta'}{\sigma})}{\fvar{\Delta}}$,
    $\calC'\theta'\sigma\submult\calC$.
\end{enumerate}
\end{lemma}
\begin{proof}
See Appendix \ref{proofsapp}.
\end{proof}
The satisfiability judgment $\asatS$ also satisfies the following properties.
\begin{lemma}
\label{foavallemma}
For any interpretations $I_1$, $I_2$, \ldots,
context $\Delta$, fact $\calC$, and substitution $\theta$,
\begin{enumerate}
  \item if $I_1\asubseteq I_2$ and $\avalS{I_1}{\Delta}{\calC}{\theta}$
    then there exist a fact $\calC'$, and substitutions $\theta'$ and $\sigma$
    s.t. $\avalS{I_2}{\Delta}{\calC'}{\theta'}$,
    $\restr{\theta}{\fvar{\Delta}}=
    \restr{(\compos{\theta'}{\sigma})}{\fvar{\Delta}}$,
    $\calC'\theta'\sigma\submult\calC\theta$;\smallskip
  \item if $I_1\asubseteq I_2\asubseteq\ldots$ and
    $\avalS{\aUn{i}{\Inf}I_i}{\Delta}{\calC}{\theta}$
    then there exist $k\in\Nat$, a fact $\calC'$, and substitutions
    $\theta'$ and $\sigma$ s.t.
    $\avalS{I_k}{\Delta}{\calC'}{\theta'}$,
    $\restr{\theta}{\fvar{\Delta}}=
    \restr{(\compos{\theta'}{\sigma})}{\fvar{\Delta}}$,
    $\calC'\theta'\sigma\submult\calC\theta$.
\end{enumerate}
\end{lemma}
\begin{proof}
See Appendix \ref{proofsapp}.
\end{proof}
We are now ready to define the abstract fixpoint operator
$\SP:\calI\rightarrow\calI$.
We will proceed in two steps. We will first define an operator working over
interpretations (i.e., elements of $\Part{\hb{}{P}}$).
With a little bit of overloading, we will call
the operator with the same name, i.e., $\SP$.
This operator should satisfy the equation
$\den{\SP(I)}=\Tp(\den{I})$ for every interpretation $I$.
This property ensures soundness and completeness of the
{\em symbolic} representation.

After defining the operator over $\Part{\hb{}{P}}$,
we will lift it to our abstract domain $\calI$ consisting
of the equivalence classes of elements of $\Part{\hb{}{P}}$ w.r.t.
the relation $\iequiv$ defined in Definition \ref{fodenotdef}.
Formally, we first introduce the following definition.
\begin{definition}[Symbolic Fixpoint Operator {\boldmath $\SP$}]
\label{fosymfixopdef}
Given an \LOf
program $P$ and an interpretation $I$, the symbolic fixpoint operator
$\SP$ is defined as follows:
$$
\begin{array}{l}
  \SP(I)\eqdef\{(\ms{H}+\calC)\,\theta\st
  (H\lollo G)\in\vrn{P},\
  \aval{I}{\SigmaP}{G}{\calC}{\theta}\}\ldotp\\
\end{array}
$$
\end{definition}
Note that the $\SP$ operator is defined using the
judgment $\asat{\SigmaP}$.

Proposition \ref{fotpspprop} states that $\SP$ is sound and complete
w.r.t $\Tp$. In order to prove this, we need to formulate Lemma
\ref{fotpsplemma} below.
\paragraph*{Notation}
Let $P$ be an \LOf program,
and $\Sigma,\Sigma_1\in\SigP$ be two signatures such that
$\Sigma_1\subseteq\Sigma$. Given a fact $\calC$, defined on $\Sigma$,
we use $\Til{\calC}{\Sigma}{\Sigma_1}$
to denote any fact which is obtained in the following way. For every constant
(eigenvariable) $c\in(\Sigma\diff\Sigma_1)$, pick a new variable in $\calV$
(not appearing in $\calC$), let it be $x_c$ (distinct variables must be
chosen for distinct eigenvariables). Now, $\Til{\calC}{\Sigma}{\Sigma_1}$
is obtained by
$\calC$ by replacing every $c\in(\Sigma\diff\Sigma_1)$ with $x_c$.
For instance, if $\calC=\{p(x,f(c)),q(y,d)\}$,
with $c\in(\Sigma\diff\Sigma_1)$ and $d\in\Sigma_1$, we have that
$\Til{\calC}{\Sigma}{\Sigma_1}=\{p(x,f(x_c)),q(y,d)\}$.

Given a context (multiset of goals) $\Delta$, defined on $\Sigma$,
we define $\Til{\Delta}{\Sigma}{\Sigma_1}$ in the same way.
Similarly, given a substitution $\theta$, defined on $\Sigma$, we use the
notation $\Til{\theta}{\Sigma}{\Sigma_1}$
to denote the substitution obtained from $\theta$
by replacing every $c\in(\Sigma\diff\Sigma_1)$ with a new variable $x_c$
in every binding of $\theta$.
For instance, if $\theta=[\substbind{u}{p(x,f(c))},\substbind{v}{q(y,d)}]$,
with $c\in(\Sigma\diff\Sigma_1)$ and $d\in\Sigma_1$, we have that
$\Til{\theta}{\Sigma}{\Sigma_1}=
[\substbind{u}{p(x,f(x_c))},\substbind{v}{q(y,d)}]$.

Using the notation
$\val{\den{I}}{\Sigma_1}{\Til{\Delta}{\Sigma}{\Sigma_1}}
{\Til{\calC}{\Sigma}{\Sigma_1}}$ we mean the judgment obtained
by replacing every $c\in(\Sigma\diff\Sigma_1)$ with $x_c$ simultaneously
in $\Delta$ and $\calC$. Newly introduced variables must not appear in
$\Delta$, $\calC$, or $I$.

When $\Sigma$ and $\Sigma_1$ are clear from the context, we simply write
$\til{\calC}$, $\til{\Delta}$, and $\til{\theta}$ for
$\Til{\calC}{\Sigma}{\Sigma_1}$, $\Til{\Delta}{\Sigma}{\Sigma_1}$, and
$\Til{\theta}{\Sigma}{\Sigma_1}$.

Finally, we use $\Xis{\Sigma_1}{\Sigma}$ (or simply $\xis$ if it is not
ambiguous) to denote the substitution which maps every
variable $x_c$ back to $c$ (for every $c\in(\Sigma\diff\Sigma_1)$),
i.e., consisting of all bindings of
the form $\substbind{x_c}{c}$ for every $c\in\Sigma\diff\Sigma_1$.
Clearly, we have that $\til{F}\xis=F$, for
any fact or context $F$, and
$\compos{\til{\theta}}{\xis}=\theta$ for any substitution $\theta$.

Note that, by definition,
$\Til{\calC}{\Sigma}{\Sigma_1}$ and $\Til{\Delta}{\Sigma}{\Sigma_1}$
are defined on $\Sigma_1$, while $\Xis{\Sigma_1}{\Sigma}$ is defined on
$\Sigma$.

\begin{lemma}
\label{fotpsplemma}
Let $P$ be an \LOf program, $I$ an interpretation,
and $\Sigma,\Sigma_1\in\SigP$ two signatures, with $\Sigma_1\subseteq\Sigma$.
\begin{enumerate}
  \item If $\aval{I}{\Sigma_1}{\Delta}{\calC}{\theta}$ then
    $\avalS{I}{\Delta}{\calC}{\theta}$;
  \item If $\valS{\den{I}}{\Delta}{\calC}$ then
    $\val{\den{I}}{\Sigma_1}{\Til{\Delta}{\Sigma}{\Sigma_1}}
     {\Til{\calC}{\Sigma}{\Sigma_1}}$.
\end{enumerate}
\end{lemma}
\begin{proof}
See Appendix \ref{proofsapp}.
\end{proof}
\begin{proposition}
\label{fotpspprop}
For every \LOf program $P$ and interpretation $I$,
$\den{\SP(I)}=\Tp(\den{I})$.
\end{proposition}
\begin{proof}
$\den{\SP(I)}\subseteq\Tp(\den{I})$.
\vsep
We prove that for every $\Sigma\in\SigP$,
$\den{\SP(I)}_\Sigma\subseteq\Tp(\den{I})_\Sigma$.
Assume $(\ms{H}+\calC)\theta\in\SP(I)$, with $H\lollo G$ a variant of a clause
in $P$ and $\aval{I}{\SigmaP}{G}{\calC}{\theta}$. Assume also that
$\calA=((\ms{H}+\calC)\theta+\calD)\theta'\in\InstS{\UpS{\SP(I)}}=
\den{Sp(I)}_\Sigma$.
We have that $\aval{I}{\SigmaP}{G}{\calC}{\theta}$ implies
$\avalS{I}{G}{\calC}{\theta}$ by item $i$ of Lemma \ref{fotpsplemma} (remember that
$\SigmaP\subseteq\Sigma$).
Therefore, by item $i$ of Lemma \ref{fovalavalrellemma}, we get
$\val{\den{I}}{\Sigma}{G\theta\theta'}{\calC'\theta'}$ for any fact
$\calC'\supmult\calC\theta$. Taking $\calC'=\calC\theta+\calD$, it follows that
$\val{\den{I}}{\Sigma}{G\theta\theta'}{\calC\theta\theta'+\calD\theta'}$.
Therefore, by definition of $\Tp$, we have
$\ms{H}\theta\theta'+\calC\theta\theta'+\calD\theta'\in(\Tp(\den{I}))_\Sigma$,
i.e., $\calA\in(\Tp(\den{I}))_\Sigma$.
\vsep
$\Tp(\den{I})\subseteq\den{\SP(I)}$.
\vsep
We prove that for every $\Sigma\in\SigP$,
$Tp(\den{I})_\Sigma\subseteq\den{\SP(I)}_\Sigma$.
Assume $\calA\in(\Tp(\den{I}))_\Sigma$. By definition of $\Tp$,
there exist a variant of a clause $H\lollo G$ in $P$,
a fact $\calC$ and a substitution $\theta$ (defined over $\Sigma$) s.t.
$\calA=\ms{H}\theta+\calC$ and $\val{\den{I}}{\Sigma}{G\theta}{\calC}$.
\vsep
By item $ii$ of Lemma \ref{fotpsplemma}
we have that $\valS{\den{I}}{G\theta}{\calC}$
implies $\val{\den{I}}{\SigmaP}{\til{G\theta}}{\til{\calC}}$
(hereafter, we use the notation $\til{\cdot}$ for
$\Til{\cdot}{\Sigma}{\SigmaP}$).
From $H\lollo G$ in $P$, we know that $G$ is defined on $\SigmaP$. It follows
easily that $\til{G\theta}=G\til{\theta}$, so that
$\val{\den{I}}{\SigmaP}{G\til{\theta}}{\til{\calC}}$.
By item $ii$ of Lemma \ref{fovalavalrellemma}, there exist a fact $\calC'$, and
substitutions $\theta'$ and $\sigma$ (defined over $\SigmaP$) s.t.
$\aval{I}{\SigmaP}{G}{\calC'}{\theta'}$,
$\restr{\til{\theta}}{\fvar{G}}=\restr{(\compos{\theta'}{\sigma})}{\fvar{G}}$,
and $\calC'\theta'\sigma\submult\til{\calC}$.
\vsep
By definition of $\SP$, we have $(\ms{H}+\calC')\theta'\in\SP(I)$.
\vsep
Now, $\calA=\ms{H}\theta+\calC=\ms{H}\til{\theta}\xis+\til{\calC}\xis=$
(note that by hypothesis $\compos{\theta'}{\sigma}$ and $\til{\theta}$ coincide
for variables in $G$, and are not defined on variables in $H$ which do not
appear in $G$ because $H\lollo G$ is a variant)
$\ms{H}\theta'\sigma\xis+\til{\calC}\xis\supmult$
$\ms{H}\theta'\sigma\xis+\calC'\theta'\sigma\xis=$
$((\ms{H}+\calC')\theta')\sigma\xis\in$
$\den{(\ms{H}+\calC')\theta'}_\Sigma\subseteq\den{\SP(I)}_\Sigma$.
\end{proof}
The following corollary holds.
\begin{corollary}
\label{equivcor}
For every \LOf
program $P$ and interpretations $I$ and $J$, if $I\iequiv J$ then
$\SP(I)\iequiv\SP(J)$.
\end{corollary}
\begin{proof}
If $I\iequiv J$, i.e., $\den{I}=\den{J}$, we have that
$\Tp(\den{I})=\Tp(\den{J})$.
By Proposition \ref{fotpspprop}, it follows that
$\den{\SP(I)}=\den{\SP(J)}$, i.e., $\SP(I)\iequiv\SP(J)$.
\end{proof}
Corollary \ref{equivcor} allows us to safely lift the definition of $\SP$
from the lattice $\couple{\Part{\hb{}{P}}}{\subseteq}$ to
$\couple{\calI}{\asubseteq}$. Formally, we define the abstract fixpoint
operator as follows.
\begin{definition}[Abstract Fixpoint Operator {\boldmath $\SP$}]
Given an \LOf program $P$ and an equivalence class $\iclass{I}$ of $\calI$,
the abstract fixpoint operator $\SP$ is defined as follows:
  $$ \SP(\iclass{I})\eqdef\iclass{\SP(I)} $$
where $\SP(I)$ is defined in Definition \ref{fosymfixopdef}.
\end{definition}
For the sake of simplicity,
in the following we will often use $I$ to denote its class $\iclass{I}$,
and we will simply use the term {\em (abstract) interpretation} to refer to
an equivalence class, i.e., an element of $\calI$.
The abstract fixpoint operator $\SP$ satisfies the following property.
\begin{proposition}[Monotonicity and Continuity]
For every \LOf program $P$, the abstract fixpoint operator $\SP$ is
mon\-o\-ton\-ic and continuous over the lattice $\couple{\calI}{\asubseteq}$.
\end{proposition}
\begin{proof}
{\em Monotonicity}.
\nl
We prove that if $I\asubseteq J$, then $\SP(I)\asubseteq\SP(J)$, i.e.,
$\den{\SP(I)}\subseteq\den{\SP(J)}$.
To prove the latter condition, we will use the characterization given
by Proposition \ref{foasubseteqprop}.
Assume $\calA=(\ms{H}+\calC)\theta\in\SP(I)$, with $H\lollo G$ a variant of a
clause in $P$ and $\aval{I}{\SigmaP}{G}{\calC}{\theta}$.
\vsep
By item $i$ of Lemma \ref{foavallemma},
there exist a fact $\calC'$, and substitutions $\theta'$ and $\sigma$
(note that they are defined over $\SigmaP$)
s.t. $\aval{J}{\SigmaP}{G}{\calC'}{\theta'}$,
$\restr{\theta}{\fvar{G}}=\restr{(\compos{\theta'}{\sigma})}{\fvar{G}}$,
$\calC'\theta'\sigma\submult\calC\theta$.
Let $\calC\theta=\calC'\theta'\sigma+\calD$, with $\calD$ a fact defined
over $\SigmaP$.
By definition of $\SP$, $\calB=(\ms{H}+\calC')\theta'\in\SP(J)$.
\vsep
Now, $\calA=(\ms{H}+\calC)\theta=\ms{H}\theta+\calC\theta=
\ms{H}\theta'\sigma+\calC'\theta'\sigma+\calD$ (note in fact that by
hypothesis $\theta'\sigma$ and $\theta$ coincide for variables in $G$, and
are not defined on variables in $H$ which do not appear in $G$ because
$H\lollo G$ is a variant).
Therefore, we have that
$\calA=\ms{H}\theta'\sigma+\calC'\theta'\sigma+\calD=\calB\sigma+\calD$.
\vsep
{\em Continuity}.
\nl
We show that $\SP$ is finitary, i.e.,
if $I_1\asubseteq I_2\asubseteq\ldots$, then
$\SP(\aUn{i}{\Inf}I_i)$ $\asubseteq$ $\aUn{i}{\Inf}\SP(I_i)$, i.e.,
$\den{\SP(\aUn{i}{\Inf}I_i)}\subseteq\den{\aUn{i}{\Inf}\SP(I_i)}$.
Again, we will use the characterization given
by Proposition \ref{foasubseteqprop}.
Assume $\calA=(\ms{H}+\calC)\theta\in\SP(\aUn{i}{\Inf}I_i)$, with
$H\lollo G$ a variant of a
clause in $P$ and $\aval{\aUn{i}{\Inf}I_1}{\SigmaP}{G}{\calC}{\theta}$.
\vsep
By item $ii$ of Lemma \ref{foavallemma}, there exist $k\in\Nat$,
a fact $\calC'$, and substitutions $\theta'$ and $\sigma$
(note that they are defined over $\SigmaP$)
s.t. $\aval{I_k}{\SigmaP}{G}{\calC'}{\theta'}$,
$\restr{\theta}{\fvar{G}}=\restr{(\compos{\theta'}{\sigma})}{\fvar{G}}$,
$\calC'\theta'\sigma\submult\calC\theta$.
Let $\calC\theta=\calC'\theta'\sigma+\calD$, with $\calD$ a fact defined
over $\SigmaP$.
By definition of $\SP$, $\calB=(\ms{H}+\calC')\theta'\in\SP(I_k)$.
\vsep
Exactly as above, we prove that
$\calA=(\ms{H}+\calC)\theta=\ms{H}\theta'\sigma+\calC'\theta'\sigma+\calD=
\calB\sigma+\calD$.
\end{proof}
\begin{corollary}
\label{fofixpoint_equivalence}
For every \LOf program $P$, $\den{\lfp{\SP}}=\lfp{\Tp}$.
\end{corollary}
Let $\SymbF{P}=\lfp{\SP}$, then we have the following main theorem.
\begin{theorem}[Soundness and Completeness]
For every \LOf program $P$, $\osP$ $=$ $\fsP$ $=$ $\den{\SymbF{P}}_\SigmaP$.
\end{theorem}
\begin{proof}
From Theorem \ref{fofixpoint_operational} and Corollary
\ref{fofixpoint_equivalence}.
\end{proof}
The previous results give us an algorithm to compute the operational and
fixpoint semantics of a program $P$ via the fixpoint operator $\SP$.
\begin{example}
\label{foSPex}
Let us consider a signature with a constant symbol $a$,
a function symbol $f$ and predicate symbols
$p,q,r,s$. Let $\calV$ be a denumerable set of variables, and
$u,v,w,\ldots\in\calV$.
Let us consider the program $P$ given below.
$$\begin{array}{l}
  1\ldotp\ \ r(w)\lollo q(f(w))\\
  [\smallskipamount]
  2\ldotp\ \ s(z)\lollo\foralx{p(f(x))}\\
  [\smallskipamount]
  3\ldotp\ \ \anti\lollo q(u)\with r(v)\\
  [\smallskipamount]
  4\ldotp\ \ p(x)\para q(x)\lollo\all\\
\end{array}$$
From clause 4, and using the first rule for $\asat{\SigmaP}$, we
get $\SP(\eset)=\iclass{\{\{p(x),q(x)\}\}}$. For simplicity, we omit the
class notation, and we write
  $$ \isp{1}=\SP(\eset)=\{\{p(x),q(x)\}\}\ldotp $$
We can now apply the remaining clauses to the element
$I=\{p(x),q(x)\}$ (remember that $\SP(\iclass{I})=\iclass{\SP(I)}$).
From the first clause (see Example \ref{foasatex}) we have
$\aval{I}{\SigmaP}{q(f(w'))}{p(x')}{[\substbind{x'}{f(w')}]}$.
It follows that $(r(w'),p(x'))[\substbind{x'}{f(w')}]=
r(w'),p(f(w'))\in\isp{2}$. As the reader can verify (see discussion
in Example \ref{foasatex}), clause 2 does not yield any further element,
and the same holds for clause 3, therefore (changing $w'$ into
$y$ for convenience)
 $$ \isp{2}=\{\{p(x),q(x)\},\{r(y),p(f(y))\}\}\ldotp $$
Now, we can apply clause 3 to the elements in $\isp{2}$. According
to Example \ref{foasatex}, we have that $
\aval{I}{\SigmaP}{q(u')\with
r(v')}{p(x''')}{[\substbind{u'}{f(y')},
     \substbind{v'}{y'},\substbind{x'''}{f(y')}]}. $
Therefore we get that $(p(x'''))[\substbind{u'}{f(y')},
\substbind{v'}{y'},\substbind{x'''}{f(y')}]=p(f(y'))\in\isp{3}$.
Clause 2 cannot be applied yet, for the same reasons as above.
Also, note that the element $r(y),p(f(y))$ is now subsumed by $p(f(y'))$.
Therefore we can assume
 $$ \isp{3}=\{\{p(x),q(x)\},\{p(f(y'))\}\}\ldotp $$
Finally, we can apply clause 2 to $\isp{3}$, using the
$\forall$-rule for the $\asat{\SigmaP}$ judgment.
Take $c\not\in\SigmaP$, and consider a renaming of the last element in
$\isp{3}$, $p(f(y''))$. Consider (a renaming of) clause 2,
$s(z')\lollo\foralx{p(f(x))}$. We have that
$\aval{I}{\SigmaP,c}{p(f(c))}{\Eps}{nil}$, with $nil$ being the empty
substitution. Therefore we get that
$\aval{I}{\SigmaP}{\foralx{p(f(x))}}{\Eps}{nil}$, from which
$s(z')\in\isp{4}$.
The reader can verify that no further clauses can be applied and
that $\isp{4}$ is indeed the fixpoint of $\SP$, therefore we have that
  $$ \isp{4}=\isp{\omega}=\{\{p(x),q(x)\},\{p(f(y'))\},\{s(z')\}\}\ldotp $$
Note that $F(P)$ is defined to be $\den{\lfp{\SP}}_\SigmaP$, therefore
it includes, e.g., the elements
$s(a)$ (see Example \ref{lofex}), $p(f(f(y'')))$ and $p(f(f(y''))),q(x'')$.
\end{example}
\sectionl{Ensuring Termination}{foterminatsec}
In general the symbolic fixpoint semantics of first order LO programs is not 
computable (see also the results in \cite{CDLMS99}).
In fact, the use of first order terms can easily lead to LO programs that
encode operations over natural numbers. 
In this section, however,
we will isolate a fragment of \LOf for which termination of the
bottom-up evaluation algorithm presented in Section \ref{foeffectivesec}
is guaranteed. An application of these results will be presented in Section
\ref{examplessec}.
First of all, we will introduce some preliminary notions that we will use later on to
prove the decidability of our fragment.

\subsection{The Theory of Well Quasi-Orderings}
\label{theoryWQO}
In the following we summarize some basic definitions and results on the
theory of {\em well quasi-orderings} \cite{Hig52,Mil85,ACJT96}.
A {\em quasi-order} $\sqsubseteq$
on a set $A$ is a binary relation over $A$ which is reflexive and transitive.
In the following it will be denoted $(A,\sqsubseteq)$.
\begin{definition}[Well Quasi-Ordering]\rm
\label{wqodef}
A quasi-order $(A,\sqsubseteq)$
is a {\em well quasi-ordering} (wqo) if for each
infinite sequence $a_0~a_1~a_2~\ldots$ of elements in $A$ there exist indices
$i<j$ such that $a_j\sqsubseteq a_i$.\footnote{Note that our $\sqsubseteq$ operator
corresponds to the $\sqsupseteq$ operator of \cite{AJ01a}.
Here we adhere to the classical logic programming convention.}

\end{definition}
We have the following results, according to which
a hierarchy of well quasi-orderings can be built starting from known ones.
In the following $\widehat{r}$ will denote the set $\{1,\ldots,r\}$, $r$ being  a natural number,
and $|w|$ the length of a string $w$.
\begin{proposition}[From \cite{Hig52}]\rm
\label{wqoresults}
\begin{enumerate}
  \item If $A$ is a finite set, then $(A,=)$ is a wqo;

  \item let $(A,\sqsubseteq)$ be a wqo, and let $A^s$ denote the set of
    finite multisets over $A$. Then, $(A^s,\entailSet)$ is a
    wqo, where $\entailSet$ is the quasi-order on $A^s$ defined as follows:
    given $S=\{a_1,\ldots,a_n\}$ and $S'=\{b_1,\ldots,b_r\}$,
    $S'\entailSet S$ if and only if there exists an {\em injection}
    $h:\widehat{n}\rightarrow\widehat{r}$ such that
    $b_{h(j)}\sqsubseteq a_j$ for $1\leq j\leq n$;

   \item let $(A,\sqsubseteq)$ be a wqo, and let $A^*$ denote the set
    of finite strings over $A$. Then, $(A^*,\sqsubseteq^*)$ is a wqo, where
    $\sqsubseteq^*$ is the quasi-order on $A^*$ defined in the following way:
    $w'\sqsubseteq^*w$ if and only if there exists a strictly monotone (meaning that
    $j_1<j_2$ if and only if $h(j_1)<h(j_2)$) {\em injection}
    $h:\widehat{|w|}\rightarrow\widehat{|w'|}$ such that
    $w'(h(j))\sqsubseteq w(j)$ for $1\leq j\leq |w|$.
\end{enumerate}
\end{proposition}
We are ready now to study the class of monadic \LOf specifications.

\subsection{Monadic \LOf Specifications}
The class of specifications we are interested in consists of {\em monadic} predicates
without function symbols.
Intuitively, in this class we can represent process that carry along a single
information taken from a possibly infinite domain (universal quantification
introduces fresh names during a derivation).
\begin{definition}[Monadic \LOf Specifications]
\label{separablespecdef}
The class of monadic \LOf specifications
consists of \LOf programs built over a signature $\Sigma$
including a {\em finite} set of constant symbols $\calL$,
no function symbols, and
a {\em finite} set of predicate symbols $\calP$
with arity at most {\em one}.
\end{definition}
\begin{definition}[Monadic Multisets and Interpretations]
\label{separablemultdef}
The class of monadic multisets
consists of multisets of (non ground) atomic formulas
over a signature $\Sigma$ including a {\em finite} set of constant symbols
$\calL$, no function symbols, and a {\em finite} set of
predicate symbols $\calP$ with arity at most {\em one}.
An interpretation consisting of monadic
multisets is called a {\em monadic interpretation}.
\end{definition}
\begin{example}
Let $\Sigma$ be a signature including a constant symbols $a$, no
function symbols, and predicate symbols $p$, $q$ and $r$
(with arity one), and $s$ (with arity zero).
Let $\calV$ be a denumerable set of variables, and
$x,y,\ldots\in\calV$. Then the clause
$$
  p(x)\para q(x)\para r(x)\para s\lollo(p(x)\para p(a))\with
  \forall v_{.}r(v)
$$
is a monadic \LOf specification, and the multiset
$\{p(x),q(y),q(x),s\}$ is a monadic multiset.
\end{example}
We have the following result.
\begin{proposition}
\label{MMnFOclosedprop}
The class of monadic  multisets is closed under applications of $\SP$, i.e.,
for every interpretation $I$, if $I$ is monadic then $\SP(I)$ is monadic.
\end{proposition}
\begin{proof}
Immediate by Definition \ref{fosymfixopdef} and Definition
\ref{foasatdef}.
\end{proof}
Following Proposition \ref{foasubseteqprop}, we define the {\em entailment}
relation between multisets of (non ground) atomic formulas, denoted
$\entailm$, as follows.
For the sake of simplicity, in the rest of this section we will apply the following 
convention.
Consider a monadic multiset.
First of all, we can eliminate constant symbols by
performing the following transformation (note that there are no other
ground terms other than constants in this class).
For every atom $p(a)$, where $p$ is a predicate symbol with arity one and
$a$ is a constant symbol in $\Sigma$, we
introduce a new predicate symbol
with arity zero, let it be $p_a$, and we transform the original multiset
by substituting $p_a$ in place of $p(a)$.
The resulting set of predicate symbols is still finite
(note that the set of constant and predicate symbols of the program is finite).
It is easy to see that entailment between multisets transformed in the
above way is a {\em sufficient} condition for entailment of the original
multisets (note that the condition is not {\em necessary}, e.g., I cannot
recognize that $p(a)$ entails $p(x)$).

Without loss of generality, we assume hereafter to deal with a set of
predicate symbols with arity {\em one} (if it is not the case, we can complete
predicate with arity less than one with {\em dummy} variables)
and without constant symbols (otherwise, we operate the transformation
previously described).
\begin{definition}
\label{MCentail}
Given two multisets $\calA$ and $\calB$,
$\calA\entailm\calB$ if and only if there exist a substitution $\theta$ and a multiset
$\calC$ such that $\calA=\calB\theta+\calC$.
\end{definition}
Then, we have the following property.
\begin{proposition}
\label{MCentailissound}
If $\calA\entailm\calB$ then $\den{\calA}\subseteq\den{\calB}$.
\end{proposition}
\begin{proof}
It follows from Definition \ref{fodenotdef} and Proposition \ref{foasubseteqprop}.
\end{proof}
Let $\calM$ be a monadic multiset with variables $x_1,\ldots,x_k$.
We define $M_{i}$ as the {\em multiset} of predicate symbols
having $x_i$ as argument in $\calM$, and $S(\calM)$ as the
multiset $\{ M_{1},\ldots, M_{k}\}$. For instance, given the
monadic multiset $\calM$ defined as
$\{p(x_1),q(x_1),p(x_1),q(x_2),r(x_2),q(x_3),r(x_3)\}$,
$S(\calM)$ is the multiset consisting  of the elements
$M_1~=~ppq$, $M_2~=~qr$, and $M_3~=~qr$, i.e.,
$S(\calM)=\{ppq,qr,qr\}$ (where $ppq$ denotes the multiset with
two occurrences of $p$ and one of $q$, and so on).
\smallskip\\
Given two multisets of multisets of predicate symbols
$S~=~\{M_1,M_2,\ldots M_k\}$ and $T~=~\{N_1,N_2,\ldots,N_r\}$,
let $S\entailSet T$ if and only if there exists an {\em injective}
mapping $h$ from $\{1,\ldots,r\}$ to $\{1,\ldots,k\}$ such that
$N_{i}\submult M_{h(i)}$ for $i:1,\ldots,r$.
As an example, $\{ppp,tt, qq, rrr\}~\entailSet~\{pp, q, rr\}$ by mapping:
$pp$ into $ppp$ ($pp\submult ppp$),
$q$ into $qq$ ($q\submult qq$), and $rr$ into $rrr$ ($rr\submult rrr$).
On the contrary, $\{ppp,rr,t,qq\}\not\entailSet\{pq,q,rr\}$, in fact
there is no multiset in the set on the left hand side of the previous relation of which $pq$
is a sub-multiset.
\smallskip\\
The following property relates the quasi order $\entailSet$ and the entailment
relation $\entailm$.
\begin{lemma}\label{ent-set}
Let $\calM$ and $\calN$ be two monadic multisets.
Then $S(\calM)\entailSet S(\calN)$ implies $\calM \entailm \calN$.
\end{lemma}
\begin{proof}
Let $S(\calM)~=~\{M_1,M_2,\ldots M_k\}$ and
$S(\calN)~=~\{N_1,N_2,\ldots,N_r\}$.
Furthermore, let $h$ be the injective mapping from $\{1,\ldots,r\}$
to $\{1,\ldots,k\}$ such that $N_{i}\submult M_{h(i)}$.
By construction of $M$ and $N$, it is easy to see that
for every $i\in\{1,\ldots,r\}$ we can isolate atomic formulas
$A_{i1},\ldots,A_{iz}$ in $\calN$ (corresponding to the cluster of
variables $N_i$), where $z$ is the cardinality of $N_i$,
and atomic formulas $B_{i1},\ldots,B_{iz}$ in $\calM$ (corresponding to the
cluster of variables $M_{h(i)}$), such that
the conditions required by Definition \ref{MCentail} are satisfied.
\end{proof}
As an immediate consequence of this lemma, we obtain the following property.
\begin{proposition}
\label{MMnFOwqoprop}
The entailment relation $\entailm$ between monadic multisets
is a well-quasi-ordering.
\end{proposition}
\begin{proof}
The conclusion follows from the observations below
(in the following we denote by $\supmult$ the converse of the sub-multiset relation, i.e.,
$\calA\supmult\calB$ if and only if $\calB\submult\calA$):
\begin{itemize}
  \item[-] the $\supmult$ relation is a well quasi-ordering by Dickson's Lemma
    (which is a consequence of Proposition \ref{wqoresults}, see also \cite{Dic13}). Intuitively,
    multiset inclusion
    is equivalent to the component-wise ordering of tuples of integers denoting
    the occurrences of the finite set of predicate symbols in a multiset;
  \item[-] since $\entailSet$ is built over elements ordered with respect to
    the well quasi-ordering
    $\supmult$, $\entailSet$ is in turn a well quasi-ordering by
    item $ii$ of Proposition \ref{wqoresults};
  \item[-] as a consequence of Lemma \ref{ent-set}, $\entailSet$ being a well quasi-ordering
    implies that $\entailm$ is a well quasi-ordering.
\end{itemize}
\end{proof}
We can now formulate the following proposition, which
states that the bottom-up fixpoint semantics is
computable in finite time for monadic \LOf specifications.
This results relies on the following facts: in the case of monadic specifications,
each interpretation computed via bottom-up evaluation consists of monadic multisets, and
the entailment relation between monadic multisets is a well quasi-ordering (therefore
eventually the fixpoint computation stabilizes).
\begin{proposition}
\label{PnFOprop}
Let $P$ be a monadic \LOf specification.
Then there exists $k\in\Nat$ such that
$\SymbF{P}$ $=$ $\bigsqcup_{i=0}^k \isp{k}(\eset)$.
\end{proposition}
\begin{proof}
We first note that the denotation of a monadic interpretation $I$
is defined in terms of the denotation of its elements (monadic multisets).
Thus, a monadic interpretation $I$ represents an {\em upward closed} set w.r.t. to the ordering
$\entailm$.
Furthermore, the sequence of interpretations computed during a fixpoint computation
forms an increasing sequence with respect to their denotation.
The result follows then from Propositions \ref{MMnFOclosedprop}, \ref{MCentailissound},
\ref{MMnFOwqoprop}, and known results on well quasi-orderings which guarantee that
any infinite increasing  sequence of upward-closed sets eventually
stabilizes (see \cite{FS01}).
\end{proof}

\sectionl{An Example}{examplessec}
In this section we show how the bottom-up semantics of
Section \ref{foeffectivesec} can be applied for verifying the
test-and-lock protocol given in Section \ref{testlocksec}.
In order to run the experiments described hereafter, we have built a
prototypical verification tool
implementing the bottom-up fixpoint procedure (backward reachability algorithm)
described in Section \ref{foeffectivesec}.
Following the guidelines and programming style
described in \cite{EP91}, we have implemented an interpreter for the relevant
first order fragment of LO, enriched with the bottom-up evaluation procedure described in Section
\ref{foeffectivesec}. The verification tool has been implemented
in Standard ML.
Let us consider again the test-and-lock protocol  given in Section \ref{testlocksec}.
Using our verification tool, we can now automatically
verify the {\em mutual exclusion} property for the
protocol. The specification of unsafe states is simply as follows:
$$\begin{array}{l}
8\ldotp\ \ use(x)\para use(x)\lollo\all
\end{array}$$
Note that the test-and-lock specification can be transformed into a  {\em monadic} one.
In fact, the second argument  can be embedded into the predicate $m$ so as to
define the two predicates $m_{unlocked}$ and $m_{locked}$.
In some sense, the specification is {\em implicitly} monadic since
the second argument is defined over a finite set of states.
Therefore termination of the fixpoint computation is
{\em guaranteed} by Proposition \ref{PnFOprop}.
Running the verification algorithm, we actually find a mutual exclusion violation.
The corresponding trace is shown in Figure \ref{testlockflawfig},
where $bc^{(i^*)}$ denotes multiple applications of clause number $i$.
The problem of the above specification lies in clause 2:
$$
  2\ldotp\ \ init\lollo init\para m(x,unlocked)\\
$$
\begin{figure*}[t]
$$
   \infer[bc^{(1^*)}]{\dedloPS{init}}
  {\infer[bc^{(2^*)}]{\dedloPS{init,think,think}}
  {\infer[bc^{(4^*)}]{\dedloPS{init,think,think,m(a,unlocked),m(a,unlocked)}}
  {\infer[bc^{(6^*)}]{\dedloPS{init,wait(a),wait(a),m(a,unlocked),
                               m(a,unlocked)}}
  {\infer[bc^{(8)}]{\dedloPS{init,use(a),use(a),m(a,locked),m(a,locked)}}
  {\infer[\all_r]{\dedloPS{init,\all,m(a,locked),m(a,locked)}}{}
  }}}}}
$$
\captionl{Incorrect test-and-lock protocol: a trace violating mutual exclusion}
  {testlockflawfig}
\end{figure*}
\begin{figure*}[t]
$$\begin{array}{l}
\{init\}\\
\{use(x),\ use(x)\}\\
\{m(x,unlocked),\ use(x),\ wait(y)\}\\
\{m(x,unlocked),\ use(x),\ use(y),\ m(y,locked)\}\\
\{m(x,locked),\ use(x),\ m(y,unlocked),\ m(y,unlocked),\ think\}\\
\{m(x,unlocked),\ m(x,unlocked),\ wait(y),\ think\}\\
\{m(x,unlocked),\ m(x,unlocked),\ use(y),\ m(y,locked),\ use(z),\
  m(z,locked)\}\\
\{m(x,unlocked),\ m(x,unlocked),\ use(y),\ m(y,locked),\ wait(z)\}\\
\{wait(x),\ m(y,unlocked),\ m(y,unlocked),\ wait(z)\}\\
\{m(x,unlocked),\ m(x,unlocked),\ think,\ think\}\\
\{use(x),\ m(x,unlocked),\ think\}
\end{array}$$
\captionl{Fixpoint computed for the incorrect test-and-lock protocol}{testlockincevalfig}
\end{figure*}
\begin{figure*}[t]
$$
   \infer[bc^{(1^*)}]{\dedloPS{init}}
  {\infer[bc^{({2'}^*)}]{\dedloPS{init,think,think,think}}
  {\infer[bc^{(3)}]{\dedloPScd{init,think,think,think,m(c,unlocked),
                               m(d,unlocked)}}
  {\infer[bc^{(4)}]{\dedloPScd{think,think,think,m(c,unlocked),
                               m(d,unlocked)}}
  {\infer[bc^{(4)}]{\dedloPScd{think,think,wait(c),m(c,unlocked),
                               m(d,unlocked)}}
  {\infer[bc^{(4)}]{\dedloPScd{think,wait(d),wait(c),m(c,unlocked),
                               m(d,unlocked)}}
  {\infer[bc^{(6)}]{\dedloPScd{wait(c),wait(d),wait(c),m(c,unlocked),
                               m(d,unlocked)}}
  {\infer[bc^{(7)}]{\dedloPScd{wait(c),wait(d),use(c),m(c,locked),
                               m(d,unlocked)}}
  {\infer[bc^{(6)}]{\dedloPScd{wait(c),wait(d),think,m(c,unlocked),
                               m(d,unlocked)}}
  {\infer*[]{\dedloPScd{use(c),wait(d),think,m(c,locked),
                               m(d,unlocked)}}
  {}}}}}}}}}}
$$
\captionl{A correct version of the test-and-lock protocol: example trace}{testlocktracefig}
\end{figure*}
\begin{figure*}[t]
$$\begin{array}{l}
\{use(x),\ use(x)\}\\
\{m(x,unlocked),\ use(x),\ init\}\\
\{m(x,unlocked),\ use(x),\ wait(y)\}\\
\{m(x,unlocked),\ use(x),\ use(y),\ m(y,locked)\}\\
\{m(x,locked),\ use(x),\ m(y,unlocked),\ m(y,unlocked),\ think\}\\
\{m(x,unlocked),\ m(x,unlocked),\ wait(y),\ think\}\\
\{m(x,unlocked),\ m(x,unlocked),\ use(y),\ m(y,locked),\ use(z),\
  m(z,locked)\}\\
\{m(x,unlocked),\ m(x,unlocked),\ use(y),\ m(y,locked),\ wait(z)\}\\
\{wait(x),\ m(y,unlocked),\ m(y,unlocked),\ wait(z)\}\\
\{m(x,unlocked),\ m(x,unlocked),\ init\}\\
\{m(x,unlocked),\ m(x,unlocked),\ think,\ think\}\\
\{use(x),\ m(x,unlocked),\ think\}
\end{array}$$
\captionl{Fixpoint computed for the correct test-and-lock protocol}{testlockevalfig}
\end{figure*}
\begin{figure*}[t]
$$\begin{array}{l}
\{use(x)\ ,use(x)\}\\
\{m(x,y)\ ,m(x,z)\}\\
\{m(x,unlocked)\ ,use(x)\ ,use(y)\ ,m(y,z)\}\\
\{m(x,unlocked)\ ,use(x)\ ,wait(y)\}\\
\{m(x,unlocked)\ ,use(x)\ ,init\}\\
\{use(x)\ ,m(x,unlocked)\ ,think\}
\end{array}$$
\captionl{Fixpoint computed using invariant strengthening for the
          test-and-lock protocol}
  {testlockinvarevalfig}
\end{figure*}
In fact, using
an (externally quantified) variable $x$ does not prevent the creation of
multiple monitors for the same resource. This causes a violation of mutual
exclusion when different processes are allowed to concurrently access a given
resource by different monitors.  Figure \ref{testlockincevalfig}
(where, for readability, we re-use the same variables in
different multisets) also shows
the fixpoint computed for the incorrect version of the protocol: note that the
singleton multiset containing the atom $init$ is in the fixpoint (this amounts
to saying that there exists a state violating mutual exclusion
which is reachable from the initial configuration of the protocol).

Luckily, we can fix the above problem in a very simple way.
As we do not care about what resource identifiers actually are,
we can elegantly encode them using universal quantification in the body of
clause 2, as follows:
$$
  2'\ldotp\ \ init\lollo init\para\foralx{m(x,unlocked)}
$$
Every time
a resource is created, a new constant, acting as the corresponding identifier,
is created as well. Note that by the operational semantics of universal
quantification, {\em different} resources are assigned {\em different}
identifiers. This clearly prevents the creation of multiple monitors
for the same resource.
An example trace for the modified specification
is shown in Figure \ref{testlocktracefig} (where
$P$ is the program consisting of clauses 1, 2', 3 through 8 (see Section \ref{testlocksec})).

Now, running again our verification tool on the corrected specification
(termination is still guaranteed by Proposition \ref{PnFOprop}), with
the same set of unsafe states, we get the fixpoint shown in Figure
\ref{testlockevalfig}. The fixpoint contains 12 elements and is reached
in 7 steps. As the fixpoint does not contain $init$, mutual exclusion
is verified, {\em for any number of processes and any number of resources}.

We conclude by showing how it is possible to optimize the fixpoint computation.
Specifically, we show that it is possible to use the so called
{\em invariant strengthening} technique in order to reduce the dimension of the sets
computed during the fixpoint evaluation. Invariant strengthening consists of enlarging the
theory under consideration with new clauses (e.g., additional clauses 
representing further
unsafe states). We remark that this technique is perfectly sound, in the sense that if no
property violations are found in the extended theory, then no violations can be found in the
original one (i.e., proofs in the original theory are still proofs in the extended one).

One possibility might be to apply the so-called {\em counting abstraction},
i.e., turn the above \LOf specification into a
propositional program (i.e., a Petri net) by abstracting first order atoms into
propositional symbols (e.g., $wait(x)$ into $wait$, and so on),
and compute the structural invariants of the corresponding Petri net.
However, this strategy is not helpful in this case (no meaningful invariant
is found). We can still try some invariants using some {\em ingenuity}.
For instance, consider the following invariant:
$$
9.\ \ m(x,y)\para m(x,z)\lollo\all
$$
For what we said previously ({\em different} resources
are assigned {\em different} identifiers) this invariant must hold for our
specification. Running the verification tool on this extended specification
we get the fixpoint in Figure \ref{testlockinvarevalfig}, containing only 6
elements and converging in 4 steps.
A further optimization could be obtained by adding the invariant
$use(x)\para m(x,unlocked)\lollo\all$ (intuitively, if someone is using a given
resource, the corresponding semaphore cannot be unlocked). In this case the
computation converges immediately at the first step.

\section{Reachability and Extensions of LO}
\label{reachability}
In this paper we have focused our attention on the relationship between
provability in LO and {\em coverability} for the configuration of a concurrent system.

Following \cite{BDM02a}, in order to characterize {\em reachability} 
problems between two ``configurations'' (goal formulas)
we need an extra feature of linear logic, namely the logical constant $\one$.
Differently from clauses with $\top$, clauses of the form 
$A_1\para\ldots\para A_n\lollo\one$ make a
derivation succeed if and only if the right-hand side of the current sequent matches an
instance of $A_1\para\ldots\para A_n$, i.e., all 
resources must be used in the corresponding derivation.

Going back to the notation used in Section \ref{unified}, 
let $P$ be a set of LO rewrite rules over $\Sigma$ and
$\calV$, and $\calM,\calM'$ two multisets of ground atomic
formulas (two configurations). Furthermore, let $H$, $G$ the
(possibly empty) $\,\para$-disjunctions of ground atomic formulas
such that $\ms{H}=\calM'$ and $\ms{G}=\calM$. Then, the
provability of the sequent  $\dedLOo{P,H\lollo\one}{G}$ precisely
characterizes the {\em reachability} of configuration $\calM'$ from the
initial configuration $\calM$ via a sequence of multiset rewriting
steps defined over the theory $P$ (see \cite{BDM02a}).
Again, this is a straightforward consequence of the properties of clauses like 
$H\lollo\one$ and of the fact that, when working  with LO rewrite rules, derivations 
have no branching.
\begin{example}
\label{msrex1}
Let us go back to Example \ref{msrex} of Section \ref{unified}
(compare the definitions of the formulas $F_1$ and $F_2$ given there).
Let $F_1'$ be the formula
$$
  p(a)\para p(f(f(b)))\para q(b)\para q(b)\para q(f(f(b)))\lollo \one
$$
and $F_2'$ be the formula
$$
  p(a)\para q(b)\lollo \one
$$
and $G=p(a)\para p(b)\para q(f(b))$.
If we enrich $P$ with $F_1'$, instead of $F_1$, then we can transform the partial derivation of
Figure \ref{LOmsr} into an LO proof as shown below 
(where $\delta$ stands for the derivation fragment of Figure \ref{LOmsr}):
$$
   \infer[bc]{\delta}
  {\infer[\one_r]{\dedloPS{\one}}{}}
$$
The resulting LO proof also shows that
from the multiset $\{p(a),p(b),q(f(b))\}$ we can reach the
multiset $\{p(a),p(f(f(b))),q(b),q(b),q(f(f(b)))\}$ after a finite number of rewriting steps
defined in accordance with $P$.
Note that on the contrary (compare with Example \ref{msrex}),
if we enrich $P$ with $F_2'$, it is {\em not}
possible to turn the partial derivation
of Figure \ref{LOmsr} into an LO proof. In fact, every rewriting step will give us
larger and larger multisets and the formula $F_2'$ never becomes applicable.
\end{example}
\begin{figure*}[t]
$$
  \infer[bc^{(1)}]{\dedlo{P,D}{\Sigma}{p(f(a)),p(b),q(f(b))}}
  {\infer[bc^{(D)}]{\dedlo{P,D}{\Sigma,c}{p(f(a)),p(f(b)),q(b),q(c)}}
   {\infer[\one_r]{\dedlo{P,D}{\Sigma,c}{\one}}{}
   }
  }
$$
\captionl{Reachability as provability in \LOfws}{LOfder}
\end{figure*}
Particular attention must be paid to the constants introduced in a derivation.
They cannot be extruded from the scope of the corresponding universal quantifier.
For this reason, the formulas representing {\em target} configurations must be generalized
by introducing universally quantified variables in place of constants introduced in a
derivation.
For the sake of brevity, we will illustrate the connection between provability and
reachability in the extended setting through the following example.
\begin{example}\label{msrex2}
Let $\Sigma$ be the signature of Example \ref{msrex}.
Let $P$ consists of the clause
$$
  1\ldotp\ \ p(x)\para q(f(y))\lollo\forall w\ldotp (p(f(x))\para q(y)\para q(w))
$$
Now, let $D$ be the clause $\forall x\ldotp p(f(a))\para p(f(b))\para q(b)\para q(x)\lollo\one$,
and let $G$ be the goal $p(f(a))\para p(b)\para q(f(b))$.
The universal quantifier is used here to generalize the representation of the target
configuration. In fact, new constants will be introduced and associated to
the predicate $q$ in the derivation of the goal $G$.
As an example, a possible derivation is shown in Figure \ref{LOfder}
(where we have omitted applications of the $\para_r$ rule for simplicity).
The last backchaining step in Figure \ref{LOfder} is possible because
of the universal quantifier used in $D$.
It would not be possible to define $D$ as
$p(f(a))\para p(f(b))\para q(b)\para q(c)\lollo\one$.
In fact, the resulting initial sequent would violate the
side condition of the $\forall_r$  proof rule that requires the
freshness of the new constants introduced in a proof.
\end{example}
The extension of the fixpoint semantics presented in this paper to more 
general linear logic languages (e.g., languages that include $\one$) is a possible
future direction for our research.
\sectionl{Conclusions}{conclusions}
In this paper we have investigated the connections between techniques used
for {\em symbolic model checking} of infinite-state systems \cite{ACJT96,FS01}
and provability in fragments of linear logic \cite{AP90}.
The relationship between the two fields is illustrated in Figure \ref{relfig}.
\begin{figure*}[t]
\begin{center}
\begin{tabular}{c|c}
   {\bf Infinite State Concurrent Systems} & {\bf Linear Logic Specification}\\
                     & \\
  transition system  & LO program and proof system\\
  transition         & rule instance\\
  current state      & goal formula\\
  initial state      & initial goal\\
  single final state & axiom with $\one$\\
  upward-closed set of states & axiom with $\top$\\
  reachability       & provability\\
  $\Pre$ operator    & $\Tp$ operator\\
  $\Pres$ operator   & $\lfp{\Tp}$\\
\end{tabular}
\end{center}
\captionl{Reachability versus provability}{relfig}
\end{figure*}
From our point of view, linear logic can be used as a unifying
framework for reasoning about concurrent systems (e.g., Petri Nets,
multiset rewriting, and so on).  In \cite{BDM02a}, we have applied
algorithms previously developed for Petri Nets in order to derive
bottom-up evaluation strategies for proposition linear logic.
Conversely, in the current paper we have shown that the use of linear
logic and the related bottom-up evaluation strategies can have
interesting application for the automated verification of
infinite-state systems in which processes are described via {\em
colored} formulas.  Several applications of the ideas presented in
this paper can be found in \cite{Boz02}, and \cite{BD02a}.

Apart from verification purposes, the new fixpoint semantics can also
be useful to study new applications of linear logic programming (e.g.,
for active databases as discussed in \cite{HW98}).  For this purpose,
it might be interesting to extend the bottom-up evaluation framework
to richer linear logic languages. Possible directions of research include
languages with a richer set of connectives (e.g., Linlog \cite{And92}),
or languages with more powerful type theories (e.g., LLF \cite{CP02}).
\subsection*{Acknowledgments}
We would like to thank the anonymous reviewers of the paper 
for their helpful comments.

\appendix

\sectionl{Some Notations}{prelimsec}
\paragraph{Multisets}
A  multiset with elements in $D$ is a function $\funct{\calM}{D}{\Nat}$.
If $d\in D$ and $\calM$ is a multiset on $D$, we say
that $d\in\calM$ if and only if $\calM(d)> 0$.
For convenience, we often use the notation for sets
(allowing duplicated elements) to
indicate multisets, when no ambiguity arises from the context. For instance,
$\{a,a,b\}$, where $a,b\in D$,  denotes the multiset $\calM$ such that
$\calM(a)=2$, $\calM(b)=1$, and $\calM(d)=0$ for all $d\in D\diff\{a,b\}$.
Sometimes we simply write $a,a,b$ for $\{a,a,b\}$.
Finally, given a set $D$, $\MS{D}$ denotes the set of multisets
with elements in $D$. We define the following operations on multisets.
Let $D$ be a set, $\calM_1,\calM_2\in\MS{D}$, and $n\in\Nat$, then:
$\Eps$ is defined s.t.
    $\Eps(d)=0$ for all $d\in D$ {\rm ({\em empty multiset})};
$(\calM_1+\calM_2)(d)=\calM_1(d)+\calM_2(d)$ for all $d\in D$
    {\rm ({\em union})};
   $(\calM_1\diff\calM_2)(d)=max\{0,\calM_1(d)-\calM_2(d)\}$
    for all $d\in D$ {\rm ({\em difference})};
    $(\calM_1\intersect\calM_2)(d)=min\{\calM_1(d),\calM_2(d)\}$
    for all $d\in D$ {\rm ({\em intersection})};
    $(n\scalar\calM)(d)=n\calM(d)$ for all $d\in D$
    {\rm ({\em scalar product})};
   $\calM_1\not=\calM_2$ if and only if there exists $d\in D$ s.t.
    $\calM_1(d)\not=\calM_2(d)$ {\rm ({\em comparison})};
    $\calM_1\submult\calM_2$ if and only if $\calM_1(d)\leq\calM_2(d)$ for all
    $d\in D$ {\rm ({\em inclusion})};
    $(\mlub{\calM_1}{\calM_2})(d)=
    max\{\calM_1(d),\calM_2(d)\}$ for all $d\in D$
    {\rm ({\em merge})};
    $\card{\calM_1}=\Sigma_{d\in D}\calM_1(d)$
    {\rm ({\em cardinality})}.
We use the notation of a formal sum $\SummL{i\in I}{\calM_i}$
to denote the union
of a family of multisets $\calM_i$, with $i\in I$, $I$ being a finite set.
It turns out that $(\MS{D},\submult)$ has the structure of a
lattice (the lattice is complete provided a greatest element is added).
In particular, merge and intersection are, respectively, the least
upper bound and the greatest lower bound operators with respect to the multiset
inclusion operator $\submult$.

\paragraph{Signatures}
Given a set of formulas $P$, we denote by $\SigmaP$ the signature comprising the set
of constant, function, and predicate symbols in $P$.
We assume to have an infinite set $\calV$ of variable symbols, usually noted
$x$, $y$, $z$, etc. In order to deal with signature augmentation
(due to the presence of universal quantification over goals)
we also need an infinite set $E$ of new constants
(called {\em eigenvariables}). We denote by $\SigP$ the set of
signatures which comprise at least the symbols in $\SigmaP$ (and possibly
some eigenvariables).

$\TsigmaV$ denotes the set of {\em non ground} terms
over $\Sigma$, i.e., the set of terms built over $\Sigma\cup V$ where $V$
is a denumerable set of variables. (A non ground term {\em may} have free variables;
a {\em ground} term is also {\em non ground}).

$\AsigmaV$ denotes the set of {\em non ground} atoms over $\Sigma$, i.e., atomic formulas
built over non ground terms over $\Sigma$.

Multisets of atoms over $\AsigmaV$ are also called {\em facts} throughout the paper, and
usually noted $\calA$, $\calB$, $\calC$, $\ldots$.

\paragraph{Substitutions and Multiset Unifiers}
We inherit the usual concept of {\em substitution} (mapping from variables
to terms) from traditional logic programming.
We always consider a denumerable set of variables $\calV$, and
substitutions are usually noted $\theta$, $\sigma$, $\tau$, \ldots
We use the notation $[\substbind{x}{t},\ldots]$, where $x$ is a variable
and $t$ is a term, to denote substitution bindings, with $nil$ denoting the
empty substitution.
The application of a substitution $\theta$ to $F$, where
$F$ is a generic expression (e.g., a formula, a term, \ldots) is denoted
by $F\theta$. A substitution $\theta$ is said to be {\em grounding} for
$F$ if $F\theta$ is ground, in this case $F\theta$ is called a
{\em ground instance} of $F$.
Composition of two substitutions $\theta$ and $\sigma$ is
denoted $\compos{\theta}{\sigma}$, e.g., $F(\compos{\theta}{\sigma})$ stands
for $(F\theta)\sigma$. We indicate the domain of a substitution $\theta$
by $\Dom{\theta}$, and we say ``$\theta$ defined on a signature $\Sigma$''
meaning that $\theta$ can only map variables in $\Dom{\theta}$ to terms in
$\TsigmaX{\calV}$.
Substitutions are ordered with respect to the ordering $\leq$
defined in this way: $\theta\leq\tau$ if and only if there exists a
substitution $\sigma$ s.t. $\tau=\compos{\theta}{\sigma}$. If $\theta\leq\tau$,
$\theta$ is said to be {\em more general} than $\tau$; if $\theta\leq\tau$
and $\tau\leq\theta$, $\theta$ and $\tau$ are said to be {\em equivalent}.
Finally, $\fvar{F}$, for an expression $F$,
denotes the set of {\em free} variables of $F$, and
$\restr{\theta}{W}$, where $W\subseteq \calV$, denotes the
{\em restriction} of $\theta$ to $\Dom{\theta}\intersect W$.

We need the notion of {\em most general unifier} (mgu).
The definition of most
general unifier is somewhat delicate. In particular, different classes of
substitutions (e.g., idempotent substitutions)
have been considered for defining most general unifiers. We refer
the reader to \cite{Ede85,LMM88,Pal90} for a discussion. Most general unifiers
form a complete lattice with respect to the ordering $\leq$, provided a
greatest element is added. For our purposes, we do not choose a particular
class of most general unifiers, we only require the operation of
{\em least upper bound}
of two substitutions w.r.t $\leq$ to be defined and effective.
The least upper bound of $\theta_1$ and $\theta_2$ is indicated
$\theta_1\slub\theta_2$. We refer the reader to \cite{Pal90} for the definition of the
least upper bound. The only property which we use in this paper is that
$\theta_1\leq(\theta_1\slub\theta_2)$ and $\theta_2\leq(\theta_1\slub\theta_2)$,
for any substitutions $\theta_1$ and $\theta_2$.
We assume $\slub$ to be commutative and associative.

We need to lift the definition of {\em most general unifier} from
expressions to multisets of expressions. Namely, given two multisets
$\calA=\{a_1,\ldots,a_n\}$ and $\calB=\{b_1,\ldots,b_n\}$
(note that $\card{\calA}=\card{\calB}$), we define a most
general unifier of $\calA$ and $\calB$, written $\Mmgu{\calA}{\calB}$, to be
the most general unifier (defined in the usual way) of the two vectors of
expressions $\tuple{a_1,\ldots,a_n}$ and $\tuple{b_{i_1},\ldots,b_{i_n}}$,
where $\{i_1,\ldots,i_n\}$ is a permutation of $\{1,\ldots,n\}$.
Depending on the choice of the permutation, in general there is more
than one way to unify two given multisets (the resulting class of mgu in
general will include unifiers which are not equivalent).
We use the notation
$\theta=\Mmgu{\calA}{\calB}$ to denote any unifier which is
{\em non deterministically} picked from the set of most general unifiers of
$\calA$ and $\calB$.

\sectionl{Proofs of Some Lemmas}{proofsapp}
\paragraph{Proof of Lemma \ref{fomovelemma}}
\nl
{\em If part}.
By induction on the derivation of $\valS{I}{\Delta,\calC}{\Eps}$.
\begin{itemize}
  \item[-] If $\Delta=\all,\Delta'$, obvious;
  \item[-] if $\Delta=\calA$ and $\calA+\calC\in I_\Sigma$, then also
    $\valS{I}{\calA}{\calC}$ holds;
  \item[-] if $\Delta=\foralx{G},\Delta'$ and
    $\val{I}{\Sigma,c}{G[c/x],\Delta',\calC}{\Eps}$, with $c\not\in\Sigma$,
    then by the inductive
    hypothesis $\val{I}{\Sigma,c}{G[c/x],\Delta'}{\calC}$, which implies
    $\valS{I}{\foralx{G},\Delta'}{\calC}$;
  \item[-] if $\Delta=G_1\with G_2,\Delta'$,
    $\valS{I}{G_1,\Delta',\calC}{\Eps}$ and
    $\valS{I}{G_2,\Delta',\calC}{\Eps}$,
    by the inductive hypothesis $\valS{I}{G_1,\Delta'}{\calC}$ and
    $\valS{I}{G_2,\Delta'}{\calC}$, which implies
    $\valS{I}{G_1\with G_2,\Delta'}{\calC}$;
  \item[-] if $\Delta=G_1\para G_2,\Delta'$ or $\Delta=\anti,\Delta'$,
    the conclusion follows by a
    straightforward application of the inductive hypothesis.
\end{itemize}
{\em Only if part}.
By induction on the derivation of $\valS{I}{\Delta}{\calC}$.
\begin{itemize}
  \item[-] If $\Delta=\all,\Delta'$, obvious;
  \item[-] if $\Delta=\calA$ and $\calA+\calC\in I_\Sigma$, then also
    $\valS{I}{\calA,\calC}{\Eps}$ holds;
  \item[-] if $\Delta=\foralx{G},\Delta'$ and
    $\val{I}{\Sigma,c}{G[c/x],\Delta'}{\calC}$, with $c\not\in\Sigma$,
    then by the inductive
    hypothesis $\val{I}{\Sigma,c}{G[c/x],\Delta',\calC}{\Eps}$, which implies
    $\valS{I}{\foralx{G},\Delta',\calC}{\Eps}$;
  \item[-] if $\Delta=G_1\with G_2,\Delta'$,
    $\valS{I}{G_1,\Delta'}{\calC}$ and $\valS{I}{G_2,\Delta'}{\calC}$,
    by the inductive hypothesis $\valS{I}{G_1,\Delta',\calC}{\Eps}$ and
    $\valS{I}{G_2,\Delta',\calC}{\Eps}$, which implies
    $\valS{I}{G_1\with G_2,\Delta',\calC}{\Eps}$;
  \item[-] if $\Delta=G_1\para G_2,\Delta'$ or $\Delta=\anti,\Delta'$,
    the conclusion follows by a
    straightforward application of the inductive hypothesis.
\end{itemize}

\paragraph{Proof of Lemma \ref{fovallemma}}
\begin{enumerate}
  \item By induction on the derivation of $\valS{I_1}{\Delta}{\calC}$.
  \begin{itemize}
    \item[-] If $\Delta=\all,\Delta'$, obvious;
    \item[-] if $\Delta=\calA$ and $\calA+\calC\in (I_1)_\Sigma$,
      then $\calA+\calC\in (I_2)_\Sigma$, because $I_1\subseteq I_2$,
      therefore $\valS{I_2}{\calA}{\calC}$;
    \item[-] if $\Delta=\foralx{G},\Delta'$ and
      $\val{I_1}{\Sigma,c}{G[c/x],\Delta'}{\calC}$, with $c\not\in\Sigma$,
      then by the inductive hypothesis
      $\val{I_2}{\Sigma,c}{G[c/x],\Delta'}{\calC}$, which implies
      $\valS{I_2}{\foralx{G},\Delta'}{\calC}$;
    \item[-] if $\Delta=G_1\with G_2,\Delta'$,
      $\valS{I_1}{G_1,\Delta'}{\calC}$ and $\valS{I_1}{G_2,\Delta'}{\calC}$,
      by the inductive hypothesis
      $\valS{I_2}{G_1,\Delta'}{\calC}$ and $\valS{I_2}{G_2,\Delta'}{\calC}$,
      which implies $\valS{I_2}{G_1\with G_2,\Delta'}{\calC}$;
    \item[-] if $\Delta=G_1\para G_2,\Delta'$ or $\Delta=\anti,\Delta'$,
    the conclusion follows by a
    straightforward application of the inductive hypothesis.
  \end{itemize}
  \item By induction on the derivation of
    $\valS{\Un{i}{\Inf}I_i}{\Delta}{\calC}$.
  \begin{itemize}
    \item[-] If $\Delta=\all,\Delta'$, then for every $k\in\Nat$,
      $\valS{I_k}{\Delta}{\calC}$;
    \item[-] if $\Delta=\calA$ and $\calA+\calC\in(\Un{i}{\Inf}I_i)_\Sigma$,
      there exists $k\in\Nat$ s.t.
      $\calA+\calC\in(I_k)_\Sigma$, i.e.,
      $\valS{I_k}{\calA}{\calC}$;
    \item[-] if $\Delta=\foralx{G},\Delta'$ and
      $\val{\Un{i}{\Inf}I_i}{\Sigma,c}{G[c/x],\Delta'}{\calC}$,
      with $c\not\in\Sigma$,
      then by the inductive hypothesis there exists $k\in\Nat$ s.t.
      $\val{I_k}{\Sigma,c}{G[c/x],\Delta'}{\calC}$,
      therefore $\valS{I_k}{\foralx{G},\Delta'}{\calC}$;
    \item[-] if $\Delta=G_1\with G_2,\Delta'$,
      $\valS{\Un{i}{\Inf}I_i}{G_1,\Delta'}{\calC}$ and
      $\valS{\Un{i}{\Inf}I_i}{G_2,\Delta'}{\calC}$,
      by the inductive hypothesis there exist $k_1,k_2\in\Nat$ s.t.
      $\valS{I_{k_1}}{G_1,\Delta'}{\calC}$ and
      $\valS{I_{k_2}}{G_2,\Delta'}{\calC}$.
      By taking $k=max\{k_1,k_2\}$, by $i$ we get
      $\valS{I_{k}}{G_1,\Delta'}{\calC}$ and
      $\valS{I_{k}}{G_2,\Delta'}{\calC}$, which implies
      $\valS{I_{k}}{G_1\with G_2,\Delta'}{\calC}$;
    \item[-] if $\Delta=G_1\para G_2,\Delta'$ or $\Delta=\anti,\Delta'$,
    the conclusion follows by a
    straightforward application of the inductive hypothesis.
  \end{itemize}
\end{enumerate}

\paragraph{Proof of Lemma \ref{fovalavalrellemma}}
\begin{enumerate}
  \item By induction on the derivation of
    $\avalS{I}{\Delta}{\calC}{\theta}$.
    \begin{itemize}
      \item[-] If $\Delta=\all,\Delta'$, obvious;
      \item[-] assume $\Delta=\calA$, with $\calB\in I$ (variant),
        $\calB'\submult\calB$, $\calA'\submult\calA$, $\calC=\calB\diff\calB'$,
        and $\theta=\restr{\Mmgu{\calB'}{\calA'}}{\fvar{\calA,\calC}}$. We want
        to prove that $\valS{\den{I}}{\calA\theta\theta'}{\calC'\theta'}$
        for every substitution $\theta'$ and fact $\calC'\supmult\calC\theta$,
        i.e., $\calA\theta\theta'+\calC\theta\theta'+\calD\theta'\in
        \den{I}_\Sigma$ for every substitution $\theta'$ and
        fact $\calD$.
        \vsep
        Now, $\calA\theta\theta'+\calC\theta\theta'+\calD\theta'=
        (\calA\theta+\calC\theta+\calD)\theta'=
        (\calA'\theta+(\calA\diff\calA')\theta+(\calB\diff\calB')\theta+\calD)
        \theta'=$
        (remember that $\calB'\submult\calB$)
        $(\calA'\theta+(\calA\diff\calA')\theta+(\calB\theta\diff\calB'\theta)+
        \calD)\theta'=$
        $\calB\theta\theta'+((\calA\diff\calA')\theta\theta'+\calD\theta')\in
        \den{I}_\Sigma$;
      \item[-] if $\Delta=\foralx{G},\Delta'$ and
        $\aval{I}{\Sigma,c}{G[c/x],\Delta'}{\calC}{\theta}$,
        with $c\not\in\Sigma$, then
        by the inductive hypothesis we have that
        $$
          \val{\den{I}}{\Sigma,c}{G[c/x]\theta\theta',\Delta'\theta\theta'}
          {\calC'\theta'}
        $$
        for every substitution $\theta'$ and fact
        $\calC'\supmult\calC\theta$ (where $\theta'$ and $\calC'$ are defined
        over $\Sigma,c$).
        \vsep
        Assuming that the variable $x$ is not in the domain
        of $\theta\theta'$ (it is always possible to rename the universally
        quantified variable $x$ in $\foralx{G}$), we have that
        $\val{\den{I}}{\Sigma,c}{G\theta\theta'[c/x],\Delta'\theta\theta'}
        {\calC'\theta'}$, and, by definition of the judgment, we get
        $\valS{\den{I}}{\foralx{(G\theta\theta')},\Delta'\theta\theta'}
        {\calC'\theta'}$, i.e.,
        $\valS{\den{I}}{(\foralx{G},\Delta')\theta\theta'}{\calC'\theta'}$,
        for every substitution $\theta'$ and fact $\calC'$ defined
        over $\Sigma,c$ (and therefore also for every substitution $\theta'$
        and fact $\calC'$ defined over $\Sigma$),
        with $\calC'\supmult\calC\theta$;
      \item[-] assume $\Delta=G_1\with G_2,\Delta'$ and
        $\avalS{I}{G_1\with G_2,\Delta'}{\calC}{\theta}$. \\
    We need to prove that
        $\valS{\den{I}}{(G_1\with G_2,\Delta')\theta\theta'}{\calC'\theta'}$
        for every substitution $\theta'$ and fact $\calC'\supmult\calC\theta$,
        i.e., that
        $\valS{\den{I}}{(G_1\with G_2,\Delta')\theta\theta'}
        {\calC\theta\theta'+\calF\theta'}$ for every substitution $\theta'$ and
        fact $\calF$.
        \vsep
        By definition of $\asatS$, we have that
        there exist facts $\calC_1'\submult\calC_1$,
        $\calC_2'\submult\calC_2$ with $\card{\calC_1'}=\card{\calC_2'}$,
        and substitutions $\theta_1,\theta_2,\theta_3$ s.t.
        $$
          \theta_3=\Mmgu{\calC_1'}{\calC_2'},\ \ \
          \calC=\calC_1+(\calC_2\diff\calC_2'),\ \ \
          \theta=\restr{(\theta_1\slub\theta_2\slub\theta_3)}
          {\fvar{\Delta,\calC}},
        $$
        $$
          \avalS{I}{G_1,\Delta'}{\calC_1}{\theta_1}\hand
          \avalS{I}{G_2,\Delta'}{\calC_2}{\theta_2}.
        $$
        By the inductive hypothesis, we have that
        $$\begin{array}{l}
          \valS{\den{I}}{(G_1,\Delta')\theta_1\theta_1'}
          {\calC_1\theta_1\theta_1'+\calD_1\theta_1'}\hand\\
          \valS{\den{I}}{(G_2,\Delta')\theta_2\theta_2'}
          {\calC_2\theta_2\theta_2'+\calD_2\theta_2'}
        \end{array}$$
        for every
        substitutions $\theta_1',\theta_2'$ and facts $\calD_1,\calD_2$.
        \vsep
        By choosing $\calD_1=(\calC_2\diff\calC_2')\theta_1+\calF_1$ and
        $\calD_2=(\calC_1\diff\calC_1')\theta_2+\calF_2$, we have,
        for every substitutions $\theta_1',\theta_2'$
        and facts $\calF_1,\calF_2$,
        $$\begin{array}{l}
          \valS{\den{I}}{(G_1,\Delta')\theta_1\theta_1'}
          {(\calC_1+(\calC_2\diff\calC_2'))\theta_1\theta_1'+
          \calF_1\theta_1'},\\
          [\smallskipamount]
          \valS{\den{I}}{(G_2,\Delta')\theta_2\theta_2'}
          {(\calC_2+(\calC_1\diff\calC_1'))\theta_2\theta_2'+
          \calF_2\theta_2'}.
        \end{array}$$
        By definition of $\theta$, we have that there exist substitutions
        ${\gamma_1},{\gamma_2},{\gamma_3}$ and $\tau$ s.t.
        $$
          \tau=\compos{\theta_1}{{\gamma_1}},\ \
          \tau=\compos{\theta_2}{{\gamma_2}},\ \
          \tau=\compos{\theta_3}{{\gamma_3}},\hand
          \theta=\restr{\tau}{\fvar{\Delta,\calC}}.
        $$
        Now, let $\calF_1$ be a variant of $\calF\theta'$ with new variables,
        and define the substitution $\theta_1'$ s.t.
        $\Dom{\theta_1'}=
        \Dom{\compos{{\gamma_1}}{\theta'}}\Union\fvar{\calF_1}$
        (clearly these two latter sets are disjoint),
        $\restr{\theta_1'}{\Dom{\compos{{\gamma_1}}{\theta'}}}=
        \compos{{\gamma_1}}{\theta'}$ and $\calF_1\theta_1'=\calF\theta'$.
        Do the same for $\calF_2$, i.e., let it be another variant of
        $\calF\theta'$ with new variables, and define $\theta_2'$ in the same
        way, so that $\Dom{\theta_2'}=
        \Dom{\compos{{\gamma_2}}{\theta'}}\Union\fvar{\calF_2}$,
        $\restr{\theta_2'}{\Dom{\compos{{\gamma_2}}{\theta'}}}=
        \compos{{\gamma_2}}{\theta'}$, and $\calF_2\theta_2'=\calF\theta'$.
        \vsep
        From the definition of $\tau$ it follows that
        $(G_1,\Delta')\theta_1\theta_1'=
        (G_1,\Delta')\theta_1{\gamma_1}\theta'=(G_1,\Delta')\theta\theta'$,
        and similarly $(G_2,\Delta')\theta_2\theta_2'=
        (G_2,\Delta')\theta\theta'$.
        Also, $(\calC_1+(\calC_2\diff\calC_2'))\theta_1\theta_1'=
        \calC\theta_1\theta_1'=\calC\theta\theta'$.
        \vsep
        We also have that
        $(\calC_2+(\calC_1\diff\calC_1'))\theta_2\theta_2'=
        (\calC_2+(\calC_1\diff\calC_1'))\theta_2{\gamma_2}\theta'=
        (\calC_2+(\calC_1\diff\calC_1'))\tau\theta'=
        (\calC_2+(\calC_1\diff\calC_1'))\theta_3{\gamma_3}\theta'=$
        (remember that $\calC_1'\submult\calC_1$)
        $(\calC_2\theta_3+(\calC_1\theta_3\diff\calC_1'\theta_3))
        \gamma_3\theta'=$
        (remember that $\theta_3$ is a unifier of
        $\calC_1'$ and $\calC_2'$)
        $(\calC_2\theta_3+(\calC_1\theta_3\diff\calC_2'\theta_3))
        \gamma_3\theta'=$
        (note that $\calC_2'\theta_3=\calC_1'\theta_3\submult\calC_1\theta_3$)
        $((\calC_2\theta_3+\calC_1\theta_3)\diff\calC_2'\theta_3)
        \gamma_3\theta'=$
        (note that $\calC_2'\submult\calC_2$)
        $(\calC_1\theta_3+(\calC_2\theta_3\diff\calC_2'\theta_3))
        \gamma_3\theta'=$
        $(\calC_1+(\calC_2\diff\calC_2'))\theta_3{\gamma_3}\theta'=$
        $\calC\theta_3{\gamma_3}\theta'=\calC\theta\theta'$.
        \vsep
        By putting everything together,
        the inductive hypotheses become
        $\valS{\den{I}}{(G_1,\Delta')\theta\theta'}
        {\calC\theta\theta'+\calF\theta'}$ and
        $\valS{\den{I}}{(G_2,\Delta')\theta\theta'}
        {\calC\theta\theta'+\calF\theta'}$, from which
        the thesis follows by definition of $\satS$;
      \item[-] if $\Delta=G_1\para G_2,\Delta'$ and
        $\avalS{I}{G_1,G_2,\Delta'}{\calC}{\theta}$, then
        by the inductive hypothesis we have that
        $\valS{\den{I}}{(G_1,G_2,\Delta')\theta\theta'}{\calC'\theta'}$, for
        every substitution $\theta'$ and fact $\calC'\supmult\calC\theta$.\\
        Therefore,
        $\valS{\den{I}}{
        G_1\theta\theta',G_2\theta\theta',\Delta'\theta\theta'}
        {\calC'\theta'}$, and, by definition of the judgment, we get
        $\valS{\den{I}}{(G_1\para G_2,\Delta)\theta\theta'}{\calC'\theta'}$;
      \item[-] if $\Delta=\anti,\Delta'$, the conclusion follows by a
        straightforward application of the inductive hypothesis.
    \end{itemize}
  \item By induction on the derivation of
    $\valS{\den{I}}{\Delta\theta}{\calC}$.
    \begin{itemize}
      \item[-] If $\Delta=\all,\Delta'$, take $\calC'=\Eps$, $\theta'=nil$,
        and $\sigma=\theta$;
      \item[-] assume $\valS{\den{I}}{\calA\theta}{\calC}$ and
        $\calA\theta+\calC\in\den{I}_\Sigma=\UpS{\InstS{I}}$. Then there exist
        $\calB\in I$, a fact $\calD$, and a substitution $\tau$ (defined on
        $\Sigma$) s.t. $\calA\theta+\calC=\calB\tau+\calD$. We can safely
        assume, thanks to the substitution $\tau$, that $\calB$ is a
        {\em variant} of an element in $I$. Also, we can assume that
        $\Dom{\tau}\subseteq\fvar{\calB}$ and
        $\Dom{\theta}\intersect\Dom{\tau}=\eset$.
        \vsep
        Now, take the
        substitution $\gamma$ s.t.
        $\Dom{\gamma}=(\Dom{\theta}\intersect\fvar{\calA})\Union\Dom{\tau}$,
        $$
          \restr{\gamma}{\Dom{\theta}\intersect\fvar{\calA}}=
          \restr{\theta}{\Dom{\theta}\intersect\fvar{\calA}}\hand
          \restr{\gamma}{\Dom{\tau}}=\tau.
        $$
        We have that
        $\calA\gamma+\calC=\calB\gamma+\calD$.
        Let $\calA'\submult\calA$ and $\calB'\submult\calB$ be two
        {\em maximal} sub-multisets s.t. $\calA'\gamma=\calB'\gamma$,
        $\rho=\Mmgu{\calA'}{\calB'}$, and
        $\theta'=\restr{\rho}{\fvar{\calA}\Union\fvar{\calB\diff\calB'}}$.
        By definition of the $\asatS$ judgment, we have that
        $\avalS{I}{\calA}{\calC'}{\theta'}$, where $\calC'=\calB\diff\calB'$.
        \vsep
        As $\gamma$ is a unifier for $\calA'$,$\calB'$, while
        $\rho=\Mmgu{\calA'}{\calB'}$, we have that there exists
        a substitution $\sigma$ s.t. $\gamma=\compos{\rho}{\sigma}$.
        Therefore, $\restr{\theta}{\fvar{\calA}}=
        \restr{\gamma}{\fvar{\calA}}=
        \restr{(\compos{\rho}{\sigma})}{\fvar{\calA}}=
        \restr{(\compos{\restr{\rho}
        {(\fvar{\calA}\Union\fvar{\calB\diff\calB'})}}{\sigma})}{\fvar{\calA}}=
        \restr{(\compos{\theta'}{\sigma})}{\fvar{\calA}}$, as required.
        \vsep
        Furthermore, since $\calA\gamma+\calC=\calB\gamma+\calD$ and
        $\calA'\submult\calA$, it follows that
        $\calA'\gamma+(\calA\diff\calA')\gamma+\calC=
        \calB'\gamma+(\calB\diff\calB')\gamma+\calD$, i.e.,
        $(\calA\diff\calA')\gamma+\calC=(\calB\diff\calB')\gamma+\calD$.
        By this equality and maximality of $\calA'$ and $\calB'$, we get
        that necessarily $(\calB\diff\calB')\gamma\submult\calC$
        (otherwise, $(\calB\diff\calB')\gamma$ and
        $(\calA\diff\calA')\gamma$ would have elements in common).
        Therefore, $\calC'\theta'\sigma=(\calB\diff\calB')\theta'\sigma=
        (\calB\diff\calB')\rho\sigma=(\calB\diff\calB')\gamma\submult\calC$,
        as required;
      \item[-] if $\Delta=\foralx{G},\Delta'$ and
        $\val{\den{I}}{\Sigma,c}{(G[c/x],\Delta')\theta}{\calC}$, with
        $c\not\in\Sigma$, then by the inductive hypothesis
        there exist a fact
        $\calC'$, and substitutions $\theta'$ and $\sigma$
        (defined over $\Sigma,c$) s.t.
        $$
          \aval{I}{\Sigma,c}{G[c/x],\Delta'}{\calC'}{\theta'},
        $$
        $\restr{\theta}{\fvar{G[c/x],\Delta'}}=
        \restr{(\compos{\theta'}{\sigma})}{\fvar{G[c/x],\Delta'}}$,
        and $\calC'\theta'\sigma\submult\calC$.
        By definition of the $\asatS$ judgment, we get that
        $$
          \avalS{I}{\foralx{G},\Delta'}{\calC'}{\theta'}.
        $$
        The conclusion follows (remember that we must ensure that
        $\calC'$, $\theta'$ and $\sigma$ are
        defined over $\Sigma$) by the following crucial observations:
        \begin{itemize}
        \item[$\cdot$] $\Dom{\theta'}\subseteq(\fvar{G[c/x],\Delta'}\Union\fvar
          {\calC'})$ by Lemma \ref{fodomainlemma};
        \item[$\cdot$] $\theta'$ does not map variables in $G[c/x],\Delta'$ to
          the eigenvariable $c$. In fact we know that $\theta$ does not map
          variables in $G[c/x],\Delta'$ to $c$ (by hypothesis) and we know that
          $\restr{(\compos{\theta'}{\sigma})}{\fvar{G[c/x],\Delta'}}=
          \restr{\theta}{\fvar{G[c/x],\Delta'}}$;
        \item[$\cdot$] $\theta'$ does not map variables in $\calC'$ to $c$
          and $\calC'$ itself does not contain $c$. In fact we know that
          $\calC$ does not contain $c$ (by hypothesis) and also that
          $\calC'\theta'\sigma\submult\calC$;
        \item[$\cdot$] we can safely assume that $\Dom{\sigma}$ does not
          contain variables mapped to $c$. Intuitively, these bindings are
          useless. Formally, we can restrict
          the domain of $\sigma$ to variables that are not mapped to $c$:
          with this restriction, the equalities
          $\restr{\theta}{\fvar{G[c/x],\Delta'}}=
          \restr{(\compos{\theta'}{\sigma})}{\fvar{G[c/x],\Delta'}}$ and
          $\calC'\theta'\sigma\submult\calC$ still hold.
        \end{itemize}
      \item[-] assume $\Delta=G_1\with G_2,\Delta'$ and
        $\valS{\den{I}}{(G_1\with G_2\Delta')\theta}{\calC}$.
        We need to prove that
        there exist a fact $\calC'$ and substitutions $\theta'$ and $\sigma$
        s.t. $\avalS{I}{G_1\with G_2,\Delta'}{\calC'}{\theta'}$,
        $\restr{\theta}{\fvar{G_1,G_2,\Delta'}}=
        \restr{(\compos{\theta'}{\sigma})}{\fvar{G_1,G_2,\Delta'}}$,
        $\calC'\theta'\sigma\submult\calC$.
        By definition of $\satS$, we have that
        $$
          \valS{I}{(G_1,\Delta')\theta}{\calC}\hand
          \valS{I}{(G_2,\Delta')\theta}{\calC}.
        $$
        By the inductive hypothesis, we have that there exist
        facts $\calC_1,\calC_2$ and substitutions
        $\theta_1,\theta_2$, $\sigma_1,\sigma_2$ s.t.
        $$
          \avalS{I}{G_1,\Delta'}{\calC_1}{\theta_1}\hand
          \avalS{I}{G_2,\Delta'}{\calC_2}{\theta_2},
        $$
        $\restr{\theta}{\fvar{G_1,\Delta'}}=
        \restr{(\compos{\theta_1}{\sigma_1})}{\fvar{G_1,\Delta'}}$,
        $\restr{\theta}{\fvar{G_2,\Delta'}}=$
        $\restr{(\compos{\theta_2}{\sigma_2})}{\fvar{G_2,\Delta'}}$,
        \linebreak
        $\calC_1\theta_1\sigma_1\submult\calC$ and
        $\calC_2\theta_2\sigma_2\submult\calC$.
        \vsep
        Now, let $\calD_1\submult\calC_1$ and $\calD_2\submult\calC_2$ s.t.
        $\calD_1\theta_1\sigma_1$ = $\calD_2\theta_2\sigma_2$ =
        $\calC_1\theta_1\sigma_1\intersect\calC_2\theta_2\sigma_2$.
        Let $\tau$ be the substitution
        $\restr{(\compos{\theta_1}{\sigma_1})}
        {\fvar{G_1,\Delta',\calC_1}}\Union
        \restr{(\compos{\theta_2}{\sigma_2})}{\fvar{G_2,\Delta',\calC_2}}$;
        $\tau$ is well defined because
        $\compos{\theta_1}{\sigma_1}$ and $\compos{\theta_2}{\sigma_2}$
        both behave like $\theta$ on variables in
        $\fvar{G_1,\Delta'}\intersect\fvar{G_2,\Delta'}$,
        and $\calC_1,\calC_2$ do not have variables in common except for
        variables in $G_1,G_2,\Delta'$ (note that new variants of elements in
        $I$ are chosen every time the judgment $\asatS$ is computed).
        \vsep
        Now, $\calD_1$ and $\calD_2$ are unified by $\tau$,
        because $\calD_1\tau=\calD_1\theta_1\sigma_1=\calD_2\theta_2\sigma_2=
        \calD_2\tau$. Therefore, there exists
        $\theta_3=\Mmgu{\calD_1}{\calD_2}$ s.t. $\tau\geq\theta_3$
        ($\theta_3$ is more general than $\tau$). Also,
        $\tau\geq\theta_1\sigma_1\geq\theta_1$ and
        $\tau\geq\theta_2\sigma_2\geq\theta_2$. Therefore,
        $\tau$ is an upper bound for $\{\theta_1,\theta_2,\theta_3\}$
        and there exist
        $\theta'=\restr{(\theta_1\slub\theta_2\slub\theta_3)}
        {\fvar{G_1,G_2,\Delta',\calC}}$, and a substitution ${\gamma}$ s.t.
        $\tau=\compos{\theta'}{{\gamma}}$.
        Now we can apply the definition of $\asatS$ (rule for $\with$) and
        we get that
        $$
          \avalS{I}{G_1\with G_2,\Delta'}{\calC'}{\theta'},
        $$
        where $\calC'=\calC_1+(\calC_2\diff\calD_2)$.
        Letting $\sigma={\gamma}$, we can prove the thesis.
        \vsep
        First of all, since
        $\compos{\theta'}{\sigma}=\compos{\theta'}{\gamma}=\tau$, and by
        definition of $\tau$, we have that
        $\restr{\theta}{\fvar{G_1,G_2,\Delta'}}=
        \restr{(\compos{\theta'}{\sigma})}{\fvar{G_1,G_2,\Delta'}}$.
        It remains to prove that $\calC'\theta'\sigma\submult\calC$ holds.
        Now, we have $\calC'\theta'\sigma$ = $\calC'\tau$ =
        $\calC_1\tau+\calC_2\tau\diff\calD_2\tau$ =
        $\calC_1\tau+\calC_2\tau\diff\calD_2\theta_2\sigma_2$ =
        $\calC_1\tau+\calC_2\tau\diff
        (\calC_1\theta_1\sigma_1\intersect\calC_2\theta_2\sigma_2)$ =
        $\calC_1\tau+\calC_2\tau\diff(\calC_1\tau\intersect\calC_2\tau)$
        $\submult$ $\calC$.
        The last passage holds because $\calC_1\tau\submult\calC$ and
        $\calC_2\tau\submult\calC$ (by definition of $\tau$ and by the inductive
        hypothesis) and relies on the following property of multisets:
        $\calA\submult\calD$ and $\calB\submult\calD$ implies
        $\calA+\calB\diff(\calA\intersect\calB)\submult\calD$;
      \item[-] if $\Delta=G_1\para G_2,\Delta'$ of $\Delta=\anti,\Delta'$,
        the conclusion follows by a
        straightforward application of the inductive hypothesis.
    \end{itemize}
\end{enumerate}

\paragraph{Proof of Lemma \ref{foavallemma}}
\begin{enumerate}
  \item Assume $\avalS{I_1}{\Delta}{\calC}{\theta}$ and
    $I_1\asubseteq I_2$.
    By item $i$ of Lemma \ref{fovalavalrellemma},
    $\valS{\den{I_1}}{\Delta\theta}{\calC\theta}$.
    By item $i$ of Lemma \ref{fovallemma},
    $\valS{\den{I_2}}{\Delta\theta}{\calC\theta}$.
    The conclusion then follows from item $ii$ of Lemma \ref{fovalavalrellemma};
  \item Assume $\avalS{\aUn{i}{\Inf}I_i}{\Delta}{\calC}{\theta}$ and
    $I_1\asubseteq I_2\asubseteq\ldots$.
    By item $i$ of Lemma \ref{fovalavalrellemma},
    $\valS{\den{\aUn{i}{\Inf}I_i}}{\Delta\theta}{\calC\theta}$, i.e.,
    as it can be readily verified from Definition \ref{fodenotdef} and
    Definition \ref{foadomaindef},
    $\valS{\Un{i}{\Inf}\den{I_i}}{\Delta\theta}{\calC\theta}$.
    By item $ii$ of Lemma \ref{fovallemma}, there exists $k\in\Nat$ s.t.
    $\valS{\den{I_k}}{\Delta\theta}{\calC\theta}$.
    The conclusion then follows from item $ii$ of Lemma \ref{fovalavalrellemma}.
\end{enumerate}

\paragraph{Proof of Lemma \ref{fotpsplemma}}
\begin{enumerate}
\item By simple induction on the derivation of
  $\aval{I}{\Sigma_1}{\Delta}{\calC}{\theta}$.
\item By induction on the derivation of $\valS{\den{I}}{\Delta}{\calC}$.
\begin{itemize}
  \item[-] If $\valS{\den{I}}{\all,\Delta}{\calC}$, immediate;
  \item[-] assume $\valS{\den{I}}{\calA}{\calC}$ and
    $\calA+\calC\in\den{I}_\Sigma$. It follows that there exist $\calB\in I$,
    a fact $\calD$, and a substitution $\theta$ (defined on $\Sigma$) such that
    $\calA+\calC=\calB\theta+\calD$. Note that $\calB$ is defined on $\SigmaP$
    by definition of (abstract) interpretation.
    \vsep
    Now, $\til{\calA}+\til{\calC}=\til{\calA+\calC}=$
    $\til{\calB\theta+\calD}=\til{\calB\theta}+\til{\calD}=$
    (remember that $\calB$ is defined on $\SigmaP\subseteq\Sigma_1$)
    $\calB\til{\theta}+\til{\calD}$.
    We can conclude that $\til{\calA}+\til{\calC}\in\den{I}_{\Sigma_1}$
    (note that $\calB\in I$ and $\til{\theta}$, $\til{\calD}$ are defined on
    $\Sigma_1$), it follows that
    $\val{\den{I}}{\Sigma_1}{\til{\calA}}{\til{\calC}}$;
  \item[-] assume $\valS{\den{I}}{\foralx{G},\Delta}{\calC}$ and
    $\val{\den{I}}{\Sigma,c}{G[c/x],\Delta}{\calC}$, with $c\not\in\Sigma$.
    From $\Sigma_1\subseteq\Sigma$ we get $\Sigma_1,c\subseteq\Sigma,c$,
    therefore we can apply the inductive hypothesis.
    It follows that
    $\val{\den{I}}{\Sigma_1,c}{\til{G[c/x],\Delta}}{\til{\calC}}$ if and only if
    $\val{\den{I}}{\Sigma_1,c}{\til{G[c/x]},\til{\Delta}}{\til{\calC}}$ if and only if
    (remember that $c\not\in\Sigma\diff\Sigma_1$ because $c\not\in\Sigma$)
    $\val{\den{I}}{\Sigma_1,c}{\til{G}[c/x],\til{\Delta}}{\til{\calC}}$.
    By definition of $\satP$ (remember that $c\not\in\Sigma$ implies
    $c\not\in\Sigma_1$), we get
    $\val{\den{I}}{\Sigma_1}{\foralx{\til{G}},\til{\Delta}}{\til{\calC}}$ if and only if
    $\val{\den{I}}{\Sigma_1}{\til{\foralx{G},\Delta}}{\til{\calC}}$
    (we assume $x$ to be disjoint with the variables introduced by the
    $\til{\cdot}$ construction);
  \item[-] the remaining cases follow by a straightforward application of the
    inductive hypothesis.
\end{itemize}
\end{enumerate}

\bibliographystyle{acmtrans}
\bibliography{biblio}

\end{document}